\documentclass[fleqn,usenatbib]{mnras}

\usepackage[T1]{fontenc}
\usepackage{ae,aecompl}

\usepackage{graphicx}	
\usepackage{amsmath}	
\usepackage{amssymb}	
\usepackage{pdflscape}
\usepackage{times,txfonts}
\usepackage{hyperref}

\def\red{}

\title[Asteroseismology in the \emph{TESS}-CVZ]{Prospects for Galactic and stellar astrophysics with asteroseismology of giant stars in the \emph{TESS} Continuous Viewing Zones and beyond}

\author[J. T. Mackereth et al.]{
J. Ted Mackereth$^{1,2,3,4}$\thanks{E-mail: tedmackereth@cita.utoronto.ca}\thanks{NSERC Banting Fellow}, Andrea Miglio$^{1}$, Yvonne Elsworth$^{1}$, Benoit Mosser$^{5}$,\newauthor Savita Mathur$^{6,7}$, Rafael A. Garcia$^{8,9}$, Domenico Nardiello$^{10,11}$, Oliver J. Hall$^{1}$,\newauthor  Mathieu Vrard$^{12}$, Warrick H. Ball$^{1}$, Sarbani Basu$^{13}$, Rachael L. Beaton$^{14,15}$\thanks{NASA Hubble Fellow}\thanks{Carnegie-Princeton Fellow},\newauthor Paul G. Beck$^{16,6}$, Maria Bergemann$^{17}$, Diego Bossini$^{18}$, Luca Casagrande$^{19}$,\newauthor Tiago L. Campante$^{18,20}$, William J. Chaplin$^{1}$, Cristina Chiappini$^{21}$, L\'eo Girardi$^{11}$,\newauthor Andreas Christ S{\o}lvsten J{\o}rgensen$^{1}$, Saniya Khan$^{1}$, Josefina Montalb\'an$^{1}$,\newauthor Martin B. Nielsen$^{1}$,  Marc H. Pinsonneault$^{22}$, Tha\'ise S. Rodrigues$^{11}$, Aldo Serenelli$^{23,24}$,\newauthor Victor Silva Aguirre$^{25}$, Dennis Stello$^{26}$, Jamie Tayar$^{27}\dagger$,  Johanna~Teske$^{28,29}\dagger$,\newauthor Jennifer~L.~van Saders$^{27}$ and Emma Willett$^{1}$\\
Affiliations are listed at the end of the paper
}

\date{}

\pubyear{2020}

\begin{document}
\label{firstpage}
\pagerange{\pageref{firstpage}--\pageref{lastpage}}
\maketitle

\begin{abstract}

The NASA-\emph{TESS} mission presents a treasure trove for understanding the stars it observes and the Milky Way, in which they reside. We present a first look at the prospects for Galactic and stellar astrophysics by performing initial asteroseismic analyses of bright ($G < 11$) red giant stars in the \emph{TESS} Southern Continuous Viewing Zone (SCVZ). Using three independent pipelines, we detect $\nu_{\mathrm{max}}$ and $\Delta\nu$ in 41\% of the 15,405 star parent sample (6,388 stars), with consistency at a level of $\sim 2\%$ in $\nu_{\mathrm{max}}$ and $\sim 5\%$ in $\Delta\nu$. 
 Based on this, we predict that seismology will be attainable for $\sim 3\times10^{5}$ giants across the whole sky and at least $10^{4}$ giants with $\geq1$ year of observations in the \emph{TESS}-CVZs, subject to improvements in analysis and data reduction techniques. The best quality \emph{TESS}-CVZ data, for 5,574 stars where pipelines returned consistent results, provide high quality power spectra across a number of stellar evolutionary states. This makes possible studies of, for example, the Asymptotic Giant Branch bump (AGBb). Furthermore, we demonstrate that mixed $\ell=1$ modes and rotational splitting are cleanly observed in the 1-year data set. By combining \emph{TESS}-CVZ data with \emph{TESS}-HERMES, \emph{SkyMapper}, APOGEE and \emph{Gaia} we demonstrate its strong potential for Galactic archaeology studies, providing good age precision and accuracy that reproduces well the age of high $\mathrm{[\alpha/Fe]}$ stars and relationships between mass and kinematics from previous studies based on e.g. \emph{Kepler}. Better quality astrometry and simpler target selection than the \emph{Kepler} sample makes this data ideal for studies of the local star formation history and evolution of the Galactic disc. These results provide a strong case for detailed spectroscopic follow-up in the CVZs to complement that which has been (or will be) collected by current surveys. 

\end{abstract}




\section{Introduction} \label{sec:intro}

Asteroseismology, the study of stellar oscillations, made possible through space based, long duration photometry of stars in missions such as  CoRoT \citep{2006cosp...36.3749B,2009A&A...506..411A}, \emph{Kepler} \citep{2010Sci...327..977B} and K2 \citep{2014PASP..126..398H} has brought about a paradigm shift in our understanding of stellar structure and evolution. Our improved understanding of stellar interiors, driven by these missions, has led in-turn to a step-change in precision on the estimates of stellar parameters such as mass, radius and age that can be ascertained from our analysis of observed oscillations and their comparison with detailed stellar modelling.
 In turn, asteroseismology provides an ideal means by which to improve and constrain such stellar models \citep[for reviews, see, e.g.][]{2013ARA&A..51..353C,2019arXiv191212300A}.
 
Stellar ages are an important aspect of the endeavour toward understanding the formation and evolution of the Milky Way, providing all important chronological context to these studies. Asteroseismic ages have already proven extremely useful in understanding aspects of the formation and evolution of the Milky Way disc \citep{2016arXiv160804951A,2018MNRAS.475.5487S,2020arXiv200414806M} and more recently, the halo \citep{2020NatAs...4..382C,2020arXiv200601783M}. Combining asteroseismic constraints with other observational methods, such as near infra-red (NIR) spectroscopy \citep[e.g. the APOKASC catalogue:][]{2014ApJS..215...19P,2018arXiv180409983P} , have allowed for the extrapolation of asteroseismic ages onto larger samples of stars for which seismic data is not available \citep{2016MNRAS.456.3655M, 2016ApJ...823..114N,2018arXiv180803278T,2019MNRAS.484..294D,2019MNRAS.489..176M}. Extending the sample size and better measuring and understanding the stars in these vital training data will no doubt play a key role in the future of asteroseismology-driven Galactic studies. Such multi-dimensional data sets, observed by multiple surveys, also provide an ideal means by which to calibrate data between surveys.

The NASA Transiting Exoplanet Survey Satellite (\emph{TESS}) mission \citep{2015JATIS...1a4003R} was designed with a focus on the detection of nearby exoplanets \citep[e.g.][]{2015ApJ...809...77S}. However, at the end of its `prime' 2-year mission, it will have provided time series photometry of stars on an all-sky basis in both targeted, short cadence data and wide field 30-minute cadence Full Frame Images (FFI)\footnote{The FFI will be re-observed in the extended mission, at 10-minute cadence, in the second half of 2020.}. This will increase the number of stars with detectable asteroseismic oscillations by at least an order of magnitude over \emph{Kepler} and CoRoT \citep[e.g.][]{2017EPJWC.16005006T, 2019ApJS..241...12S,2019arXiv191207604S}. Indeed, the asteroseismic potential of \emph{TESS} has already been explored using early data products \citep[e.g.][]{2019AJ....157..245H,2019ApJ...885...31C,2020NatAs...4..382C} . The first two years of \emph{TESS} all-sky observations were taken in 27 day sectors that overlap at the ecliptic poles, forming what is referred to as the \emph{continuous viewing zones} (CVZ), within which a complete year of continuous data has now been gathered. Stars in the CVZs are likely to have power spectra that better sample lower frequency signals than those with single sectors of data. The data for these stars can be analysed in greater detail, offering higher fidelity insights into their interiors and generating more precise estimates of their parameters.

In this paper, we present a first look at the asteroseismic constraints which are possible for stars in the \emph{TESS}-SCVZ, based on the publicly released full frame images (FFI) of the first year of \emph{TESS} data. The SCVZ data has been fully available for over a year, providing ample time for detailed reduction of its time-series data. By selecting a sample of very bright ($G < 11$) giant stars based on \emph{Gaia} DR2 \citep{2018arXiv180409365G} and 2MASS \citep{2006AJ....131.1163S} photometry, we demonstrate that the detection of stellar oscillations in the CVZ data across many evolutionary stages on the giant branch will allow detailed studies of stellar structure and evolution. This data will facilitate studies of the nearby Galactic stellar populations in age space, providing a strong justification for the necessity of gathering extended spectroscopic data for such samples.

In Section \ref{sec:data}, we present the \emph{TESS}-SCVZ bright giant sample, outlining the sample selection criteria and photometry of the FFI as well as presenting the external spectroscopic, photometric and kinematic constraints that we use to study the properties of the sample. Section \ref{sec:results} presents the results of our asteroseismic analyses. There, we discuss the seismic detection yields before showing the potential of these data for stellar and Galactic astrophysics. Finally, in Section \ref{sec:conclusions} we summarise our findings and make conclusions on the potential of the data-set and the extrapolation of these results to the all-sky sample.

\section{The \emph{TESS}-SCVZ bright giant sample} \label{sec:data}

We first describe the compilation of a catalogue of stars in and around the \emph{TESS} southern continuous viewing zone for which we aim to achieve asteroseismic constraints. To this end, we select targets from \emph{Gaia} and 2MASS, whose photometry is then extracted from the \emph{TESS} full frame images (FFI) and processed to asteroseismic power spectra. These spectra are then analysed by three independent pipelines to establish the global seismic parameters $\nu_{\mathrm{max}}$ and $\Delta\nu$. We complement this data with spectroscopic constraints which include the necessary stellar parameters to establish estimates of the stellar mass (and therefore age) using the Bayesian tool \texttt{PARAM} (described below). Furthermore, we include kinematic constraints which allow the inspection of this data in full six-dimensional phase space. The data set is described below, and the resulting catalogue (available online) in Appendix \ref{app:D}.

\subsection{The parent sample: \emph{Gaia} and 2MASS data}

\begin{figure}
    \centering
    \includegraphics[width=0.9\columnwidth]{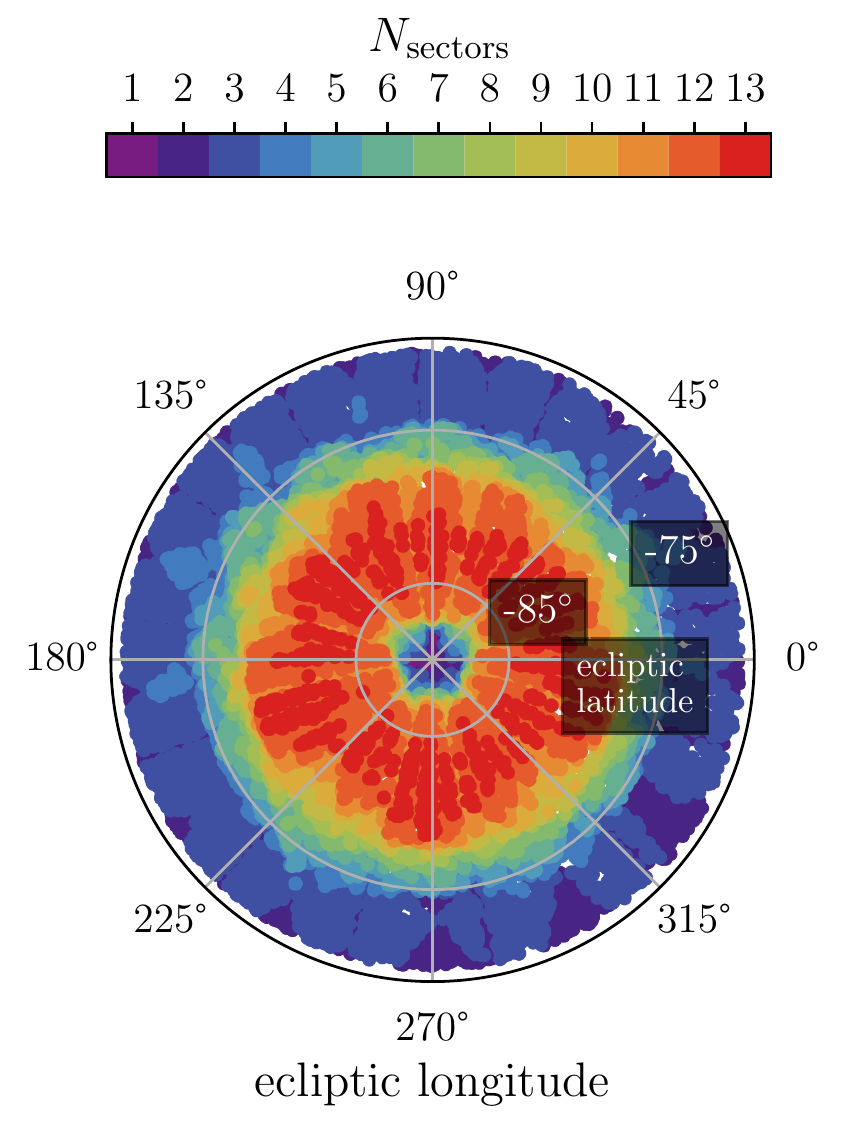}
    \caption{The \emph{TESS}-SCVZ bright giant $G < 11$ sample in polar projection, demonstrating how the number of available sectors changes with position on sky, due to the \emph{TESS} pointing scheme. Each camera has a $24^{\circ} \times 24^{\circ}$ field-of-view, providing a $\sim 24^{\circ}$ circular zone with many stars with 13 sectors of continuous observations. In this bright sample, many geometric selection effects are visible, due mainly to gaps in the camera CCDs. }
    \label{fig:polar}
\end{figure}

We compile a target list of stars in and near to the southern CVZ (SCVZ), for which \emph{TESS} observations are now complete, by making a cone-search from \emph{Gaia} DR2 \citep{2018arXiv180409365G} within $20^{\circ}$ of the southern ecliptic pole. We cross match this set with 2MASS \citep{2006AJ....131.1163S} and select stars with $G < 11$ and parallax signal-to-noise $\varpi/\delta\varpi > 20$ (i.e. uncertainties of $< 5\%$), isolating the brightest targets with the most precise parallax measurements, for which we expect the greatest yield of asteroseismic parameters. Because we select the brightest and therefore nearby stars, we are likely unaffected by the population effects from parallax SNR selection discussed by \citet{2018A&A...616A...9L}. We then select stars with $(J-K_S) > 0.5$ and $M_H < 3$ (estimating $M_H$ using only the inverted parallax and ignoring the effects of extinction - although we do account for this later), isolating a final sample of 15,405 giant stars including 3,019 with at least 12 sectors (27-day chunks) of data. The `true' CVZ is within $\sim 10^{\circ}$ of the ecliptic pole, so this sample includes 12,386 stars that have less than $12$ sectors of data. The data in this region of the sky therefore has a wide range in dwell time and observing pattern, making it ideal for tests of data products and yields all-sky. The on-sky distribution of the stars in the input catalogue is shown in polar projection in \autoref{fig:polar}. The number of sectors for which photometry was recovered is indicated by the colour of the points, demonstrating the geometric effects imposed on the data by the pointing scheme of \emph{TESS}. Inside the CVZ, there are a number of stars with less than optimal length time-series. Furthermore, the sample is clearly biased to be nearby, such that $95\%$ of targets have inverse parallaxes of $d < 1.7\ \mathrm{kpc}$. Importantly, geometric and distance selection effects may be necessary to account for in future studies which require forward modelling.

\subsection{Extraction of light curves from the \emph{TESS} FFI}

We extract photometry from the \emph{TESS} Full Frame Images (FFI) for all 15,405 stars in the target list, following the methods presented in \citet{2019MNRAS.tmp.2509N}. Briefly, this method models the \emph{TESS} PSF, accounting for spatial and temporal variations to perform photometry and neighbour subtraction (for sources between 10$^{\prime\prime}$ and $400^{\prime\prime}$ from the target star) where fields are crowded. We extract light curves from the FFI using the \texttt{img2lc} code presented in \citet{2015MNRAS.447.3536N,2016MNRAS.455.2337N}. The pipeline corrects for some systematic effects associated with the spacecraft, detector, and environment by modelling them using the co-trending basis vectors (CBVs) obtained by \citet{2020MNRAS.495.4924N}.  As an example, this allows for reconstruction of the light curves where a pointing problem caused a systematic loss of flux in sector 1. 

The resulting light curves are then post-processed using methods outlined in \citet{10.1111/j.1745-3933.2011.01042.x}. Gaps in the light curves which are larger than 3 sectors ($\sim 81$ days) are removed, while smaller gaps are in-painted. {\red We close up gaps longer than 3 sectors by concatenating the end of one sector with the beginning of the next in order to reduce the effect of the window function containing the gaps in the PSD. For modes that have a lifetime shorter than 3 sectors (~90 days), the modes have been re-excited and no effect can be seen in the PSD. For modes with longer lifetimes, a break in the phase is introduced (with the same effect as a stochastic excitation), which tends to slightly widen the peaks in the PSD of these modes (g-dominated mixed modes), while the frequencies are unchanged. Because we are only interested in the frequency of the modes and not in extracting the lifetimes, this methodology is justified in this case.} This also removes artefacts in the data resulting from outlying data points due to target drift at the start of sectors (e.g. as the spacecraft finalised its positioning). Finally, a high-pass filter at 2 days ($\sim 5\ \mathrm{\mu Hz}$) is applied to the time-series to remove any long term trends (this affects the detection of oscillations lower than $\nu \sim 10\ \mathrm{\mu Hz}$). These light curves are then transformed into power spectra in frequency space using a Lomb-Scargle periodogram estimate with an oversampling factor of 10. {\red Such an oversampling can cause small biases in the determination of seismic parameters, but this is greatly reduced for the low frequency, stochastically excited modes in giant stars, where such oversampled spectra can improve detection statistics. } 

The power spectra are analysed by three pipelines to determine the global parameters $\nu_{\mathrm{max}}$, the frequency at maximum power, and $\Delta\nu$, the mean spacing between pressure modes of the same degree $l$ at successive radial orders. The pipelines in question are presented in \citet{2009A&A...508..877M}, \citet{Elsworth_2020} and \citet{2010A&A...511A..46M}; we refer to them here as COR, BHM and A2Z, respectively \citep[similarly to, e.g.][]{2018arXiv180409983P}. {\red COR and A2Z use the power spectra for analysis while BHM uses directly the time-series data.} All the pipelines use independent approaches to determine the parameters, providing a means by which to assess the internal consistency in the results.

\subsection{Spectroscopic and photometric parameters}

In order to robustly determine useful stellar parameters such as mass and radius using the global seismic parameters, independent measures of the stellar effective temperature $T_{\mathrm{eff}}$ and optionally, as additional constraints, surface gravities $\log(g)$ and/or metallicity $Z$, are necessary. In the case of the \emph{TESS}-SCVZ, there is little publicly available spectroscopic data from which to derive these quantities (as of the preparation of this manuscript). We demonstrate below that the gathering of detailed spectroscopy for these targets will be of great utility for the community. However, for the purposes of this `proof-of-concept' study, we combine constraints on these quantities from the catalog of $T_{\mathrm{eff}}$ and $\mathrm{[Fe/H]}$ derived in \citet{2019MNRAS.482.2770C} from the SkyMapper photometric survey DR1.1 \citep{2018PASA...35...10W} with the \emph{TESS}-HERMES DR1 data \citep{2018MNRAS.473.2004S}, which provides spectroscopic constraints on $\log{(g)}$ (but not yet on $T_{\mathrm{eff}}$ or $\mathrm{[Fe/H]}$ for these bright giants). In total, we find that only 1186 of the 15405 giants have all three parameters available in these catalogues. By requiring a constraint only on $T_{\mathrm{eff}}$ \citep[e.g. as required when deriving the mass and radii from the asteroseismic scaling relations: ][]{1995A&A...293...87K} the sample size increases to 8249 stars. For the stars without spectroscopic $\log(g)$ from \emph{TESS}-HERMES, we infer a posterior on $\log(g)$ using the sample with full spectro-photometric information as a training set using the methodology described in Appendix \ref{app:A}.

We also make use of the SDSS-IV/APOGEE-2 DR16 \citep{2015arXiv150905420M,2019arXiv191202905A} spectroscopic catalogue of southern stars to perform a cross check of parameters derived using SkyMapper and \emph{TESS}-HERMES stellar parameters. APOGEE also includes detailed element abundance information derived through the application of APOGEE Stellar Parameters and Chemical Abundances Pipeline (ASPCAP: \citeauthor{2016AJ....151..144G} \citeyear{2016AJ....151..144G}) -- which uses a specifically derived linelist (\citeauthor{2015ApJS..221...24S} \citeyear{2015ApJS..221...24S}) --  to spectra of the NIR $H$-band taken using the twin (Southern) APOGEE spectrograph \citep{2019arXiv190200928W} on the 2.5m Irénée du Pont telescope at Las Cumbres Observatory \citep[LCO:][]{Bowen:73}. APOGEE spectra are then reduced and analysed using in-house pipelines \citep{2015AJ....150..173N,2020arXiv200705537J}. Red giants in the \emph{TESS}-CVZs are specifically targeted within APOGEE via external program time through the Carnegie Institution of Science (PIs: Beaton, van Saders, and Teske). Information about these extra targets and more generally about APOGEE-2 targeting is presented in Beaton et al. (2020, in prep.) and Santana et al. (2020, in prep.). Using the APOGEE abundances, we can make some initial insights into the connections between stellar age and element abundances in the Galaxy. We find that APOGEE-2 DR16 has 513 targets in common with our bright \emph{TESS}-CVZ giants. The SkyMapper $T_\mathrm{eff}$ and $\mathrm{[Fe/H]}$ generally agree well (within $1\sigma$) with those derived by APOGEE, in the cases where both surveys returned these parameters.

We use these spectroscopic and photometric stellar parameters to determine bolometric corrections in the $J$, $H$ and $K_S$ band for each of the targets represented in all the relevant data sets. We use the \texttt{bolometric-corrections} code\footnote{\url{https://github.com/casaluca/bolometric-corrections}}, which applies methods described in \citet{2014MNRAS.444..392C}. We then use this correction, in conjunction with the \emph{Gaia} DR2 parallax information and the 2MASS photometry to determine the luminosity of each of the targets. The $K_{S}$ band extinction $A_{K_S}$ is determined for each target using the \texttt{Combined19} dustmap \citep[built from the combined dustmaps of][]{2003A&A...409..205D,2006A&A...453..635M,2019ApJ...887...93G} implemented in the \texttt{mwdust}\footnote{\url{https://github.com/jobovy/mwdust}} Python package \citep{2016ApJ...818..130B}. We transform $A_{K_S}$ for the $J$ and $H$ band using the ratios determined by \citet{2005ApJ...619..931I}. We calculate the uncertainty on the luminosity by propagating those on $T_{\mathrm{eff}}$, $\log(g)$ and $\mathrm{[Fe/H]}$ into the bolometric correction. Parallax uncertainties are propagated when computing absolute magnitudes. 
We compare the photometric luminosity with that determined from seismic parameters and scaling relations in Appendix \ref{app:B}, and discuss this in relation to systematics in \emph{TESS} in Section \ref{sec:systematics}. 

The parameters are then used, in conjunction with the pipeline constraints on $\nu_{\mathrm{max}}$ and $\Delta \nu$, to determine mass, radius and age estimates for the sample via a Bayesian comparison with stellar models using the \texttt{PARAM} code \citep{2006A&A...458..609D,2014MNRAS.445.2758R,2017MNRAS.467.1433R}. We generate results separately based on the \emph{SkyMapper}/\emph{TESS}-HERMES and APOGEE DR16 parameters, including both in our catalogue. \texttt{PARAM} provides full posterior information on age and mass based on the input observables (and their uncertainties). We report here and in the catalogue the median and inter-quartile range of the posterior distributions as our final mass, radius and age constraints. 

{\red While \texttt{PARAM} provides a robust way to estimate mass and age based on the seismic and spectro-photometric constraints, it has a number of important caveats. Of course, any comparison to stellar models is subject to the limitations of the model predictions themselves. Similarly, the comparison between models and data is hampered by details such as the so-called surface effects which are not currently modelled and likely depend on stellar parameters in complex ways \citep[e.g.][]{2018A&A...620A.107M}. However, a number of tests have shown that this method provides accurate mass estimates to within a few per-cent when compared to eclipsing binaries and clusters \citep[e.g.][]{2016MNRAS.461..760M,2017MNRAS.467.1433R,2017MNRAS.472..979H,2018MNRAS.476.3729B}. The \texttt{PARAM} approach also avoids the usual problems in understanding uncertainties associated with corrections to the $\Delta\nu$ scaling relation \citep[e.g.][]{2018MNRAS.476.3729B}.}

\subsection{Stellar kinematic constraints}

We obtain kinematic constraints for the sample using astrometric parameters from the \emph{Gaia} DR2 \citep{2018arXiv180409365G} catalogue. The proper motion $\mu_{[l,b]}$ and radial velocity $v_{\mathrm{helio}}$ constraints provided by \emph{Gaia} in this bright and nearby regime are likely to be accurate, and so we apply these without correction or adjustment for zero-point offsets or biases, which are small in comparison to parameter values. The more important parallax zero-point offsets in the (now extensive) literature \citep[e.g.][]{2018A&A...616A...2L,2019MNRAS.487.3568S,2019A&A...628A..35K,2019MNRAS.486.3569H,2019arXiv190208634L,2019ApJ...878..136Z,2020MNRAS.493.4367C} range from $30 \lesssim \Delta\varpi \lesssim 60\ \mathrm{\mu as}$, agreeing that the raw \emph{Gaia} DR2 values are too small. The majority of these groups consistently find an offset in the region of $50\ \mathrm{\mu as}$. We make a simplified assessment of the parallax zero-point offset implied by our seismic results in Appendix \ref{app:C}, finding an offset of $30 \pm 2\ \mathrm{\mu as}$ for this nearby, bright sample, which we apply externally to every star in our catalogue. This implies a $\lesssim 3\%$  decrease in distance to the majority of the stars we consider. We defer a more detailed assessment of the \emph{Gaia} parallax zero-point offset using \emph{TESS} to future studies.

To propagate astrometric parameters to Galactocentric coordinates, we take 100 samples of the joint posterior of the parameters using the median, uncertainty and correlation coefficients for each. Throughout the paper, we adopt a solar position of $[R_{\odot}, z_{\odot}] = [8.125,0.02]$ kpc \citep{2018A&A...615L..15G,2018arXiv180903507B} and a corresponding velocity  $\vec{v}_{\odot} = [U,V,W] = [-11.1,245.6,7.25]\ \mathrm{km\ s^{-1}}$ based on the combined constraints of \citet{2010MNRAS.403.1829S} and the SGR A* proper motion from \citet{2018A&A...615L..15G}. We process every sample of the astrometric parameters into the left-handed Galactocentric cylindrical coordinate frame. The resulting uncertainties on $v_{R}$, $v_T$ and $v_z$ are $\sim 1\ \mathrm{km\ s^{1}}$.

 Finally, we estimate the orbital parameters $r_{\mathrm{peri}}$, $r_{\mathrm{apo}}$, $e$ and $z_{\mathrm{max}}$ for each sample of each stars phase space coordinates using the fast orbit estimation method described by \citet{2018arXiv180202592M} and implemented in \texttt{galpy} \citep{2015ApJS..216...29B}, adopting the simple \texttt{MWPotential2014} potential included there. We include these estimations in our final catalogue, reporting the median and inter-quartile range of the resulting posterior distribution of orbital parameters for each star. The median uncertainties on the final orbital parameters are less than $2\%$.


\section{Results} \label{sec:results}

\subsection{Systematics in light curves derived from the FFI}
\label{sec:systematics}

Before looking in detail at our seismic results, we remove spurious seismic detections by comparing the luminosities computed based on the seismic parameters with those from \emph{Gaia}. Stars that are erroneously assigned a low (high) $\nu_{\mathrm{max}}$ by any pipeline should have brighter (fainter) seismic luminosities relative to those derived directly from photometry, and so can be removed from further analysis. We perform this check by computing a `seismic' luminosity $L_{\mathrm{seis.}}$, via the asteroseismic scaling relation for the stellar radius $R$,
\begin{equation}
    \left(\frac{R}{R_{\odot}}\right) \simeq \left(\frac{\nu_{\mathrm{max}}}{\nu_{\mathrm{max}_{\odot}}}\right)\left(\frac{\Delta\nu}{\Delta\nu_{\odot}}\right)^{-2}\sqrt{\frac{T_{\mathrm{eff}}}{T_{\mathrm{eff}_{\odot}}}}
\end{equation}
which can then be substituted into the relationship between the luminosity and radius:
\begin{equation}
    \left(\frac{L}{L_{\odot}}\right) \simeq \left(\frac{R}{R_{\odot}}\right)^{2} \left(\frac{T_{\mathrm{eff}}}{T_{\mathrm{eff}_{\odot}}}\right)^{4}
\end{equation}
achieving a relationship between luminosity and the global seismic parameters
\begin{equation}
    \left(\frac{L}{L_{\odot}}\right) \simeq \left(\frac{\nu_{\mathrm{max}}}{\nu_{\mathrm{max}_{\odot}}}\right)^{2}\left(\frac{\Delta\nu}{\Delta\nu_{\odot}}\right)^{-4}\left(\frac{T_{\mathrm{eff}}}{T_{\mathrm{eff}_{\odot}}}\right)^{5}.
\end{equation}
For each target where a photometric luminosity $L_{\mathrm{phot.}}$ could be estimated using the spectrophotometric parameters, we take samples of the asteroseismic parameters and $T_{\mathrm{eff}}$ assuming uncorrelated Gaussian uncertainties. We then propagate these samples through the above relationship to attain $L_\mathrm{seis.}$ and it's associated uncertainty. Clearly, the precision and accuracy of $L_\mathrm{seis.}$ can be improved beyond that achievable with scaling relationships \citep[e.g.][]{2019A&A...628A..35K}, but this methodology is sufficiently accurate to detect false-positives.

We demonstrate the full comparison for each pipeline in Appendix \ref{app:B}, but briefly summarise the results here. Between $\sim 5$ to $15\%$ of targets have a seismic luminosity which is more than $3\sigma_c$ (where $\sigma_c$ is defined at the uncertainties on each measurement added in quadrature) from the photometric value. The majority of problematic cases occur in bright stars, where pipelines return erroneously high $\nu_{\mathrm{max}}$ measurements. This is expected, since the frequency resolution of \emph{TESS} is limited, making characterisation of the power spectrum more difficult at low frequencies (i.e. $\nu \lesssim 10\ \mathrm{\mu Hz}$). As an example, for 13 sectors of data, at a $\nu_{\mathrm{max}} \sim 1\ \mathrm{\mu Hz}$ the number of independent frequency bins in the spectrum around the power envelope is $\sim 20$. It should also be noted that the high-pass filtering applied to the light curves will strongly affect any possibility of detection below $\sim 10\ \mathrm{\mu Hz}$. We flag targets with $|(L_{\mathrm{phot.}}-L_{\mathrm{seis.}}) / \sigma_c| > 3$ for each pipeline in our final catalogue. Such discrepancies may also be explained by unresolved binary systems \citep{2014ApJ...784L...3M}, which have a higher than expected apparent luminosity, while the seismology represents usually just one component of the binary. The detection of such systems requires highly precise parallax measurements, such as those for these nearby targets, which have hitherto been unavailable to seismic samples (e.g. those from \emph{Kepler}). Similarly, we expect that there should be a large presence of wide binary systems within this sample \citep[such as those found in \emph{Kepler}, e.g.][]{2018MNRAS.479.4440G}, which can be useful for calibrating independent age measurement techniques \citep[e.g.][]{2012ApJ...746..102C}.

A number of targets also have apparently spurious detections at the diurnal frequency, $\nu_{\mathrm{max}} \simeq 11.57\ \mathrm{\mu Hz}$. We note that while many of these detections are consistent with that expected based on their photometric luminosities, at least some appear to be due to some spurious power excess at roughly this frequency. Extraction of the background signal from the FFI does appear to show such an excess, suggesting that this is due to some still existent issue with e.g. scattered light \citep[such issues are noted in the \emph{TESS} Data Release notes,][]{tessdrnotes}. To maximise the value of these data for asteroseismology, these systematics must be studied further and accounted for. Detailed analysis and correction for this is beyond the scope of this exploratory paper, and we simply remove problematic cases from our analysis. 

\subsection{Detection yields and seismic constraints}

\begin{figure}
    \centering
    \includegraphics[width=0.9\columnwidth]{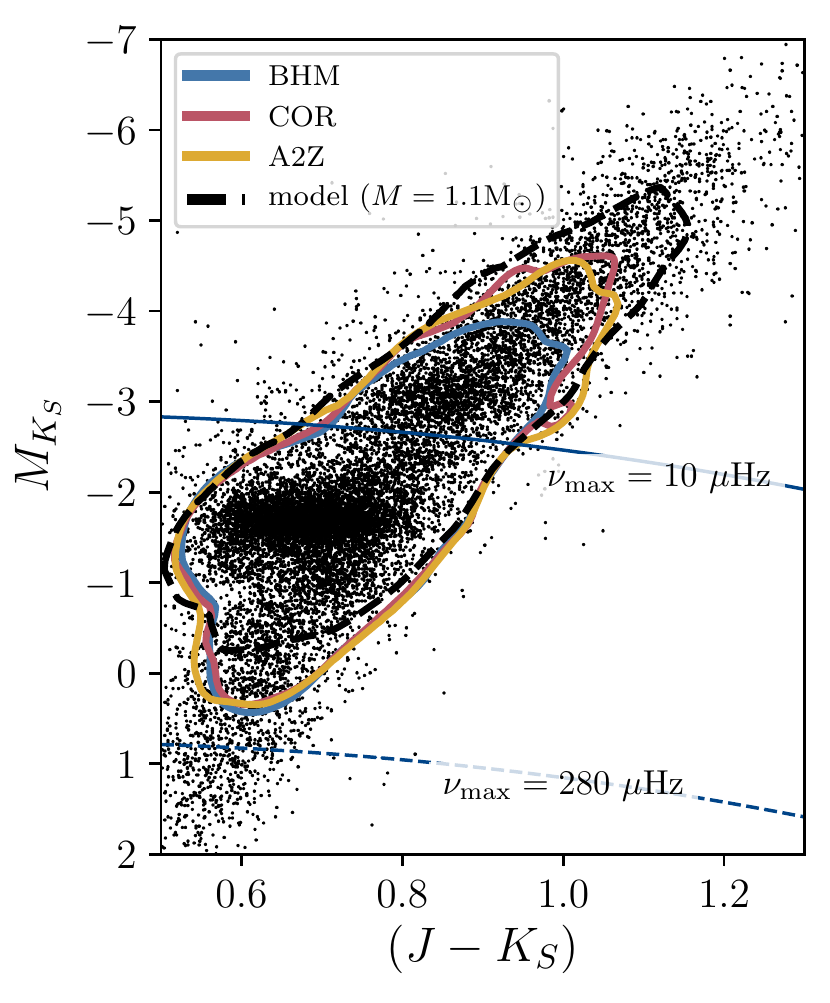}
    \caption{$M_{K_S}$-$(J-K_S)$ of the \emph{TESS}-SCVZ bright giant sample. The contours demonstrate the regions in this space that contain 90\% of the detections made by each of the pipelines (confirmed using photometric luminosities) and a model for the detection yield based on \citet{2011ApJ...732...54C} and \citet{2019ApJS..241...12S} (described further in the text). The boundaries where $\nu_{\mathrm{max}}$ is equal to $10$ and $280\ \mathrm{\mu Hz}$ for a $1.1\ \mathrm{M_{\odot}}$ star. $10\ \mathrm{\mu Hz}$ is likely the limit at which our light curve processing might affect detections and $280\ \mathrm{\mu Hz}$ is roughly the Nyquist frequency of TESS. The \emph{TESS}-SCVZ CMD has a number of important features, such as a prominent red clump (RC) at $M_{K_S} \simeq -1.5$, a clear red giant branch (RGB) extending over the full range in $M_{K_S}$ and the asymptotic giant branch bump at $M_{K_S} \simeq -3$. Each pipeline makes asteroseismic detections across all of these evolutionary stages.}
    \label{fig:cmd}
\end{figure}

\begin{figure*}
    \centering
    \includegraphics[width=0.9\textwidth]{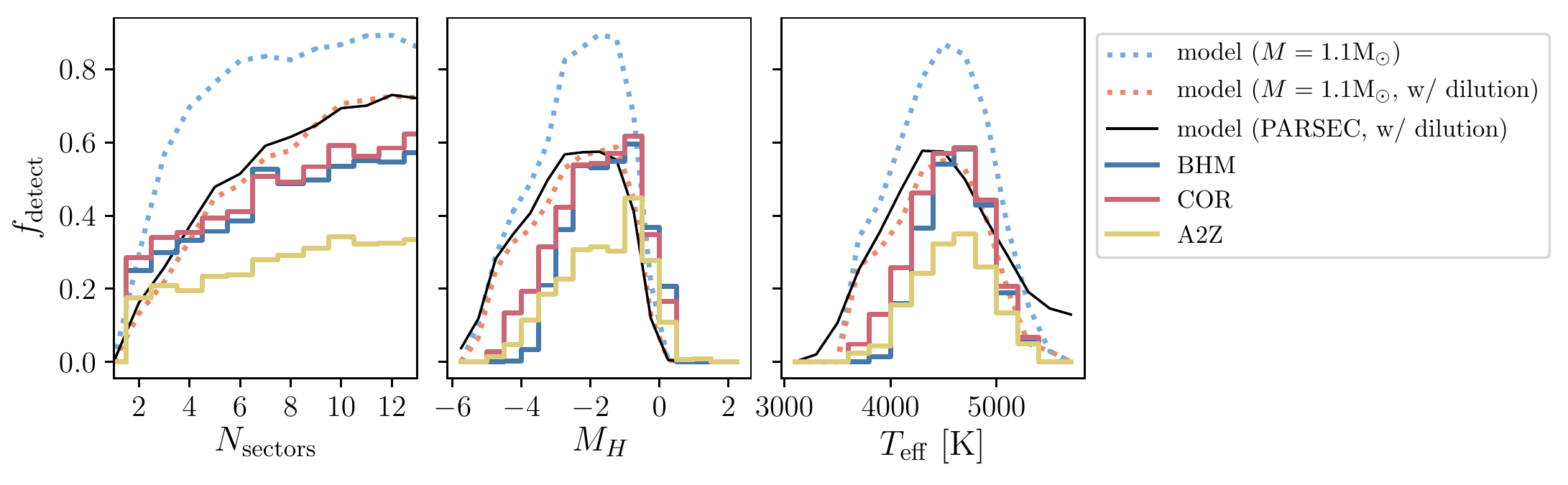}
    \caption{The detection yield $f_\mathrm{detect}$ as a function of the number of sectors $N_\mathrm{sectors}$, absolute $H$-band magnitude $M_H$ and effective temperature $T_{\mathrm{eff}}$ from SkyMapper. The coloured histograms show $f_{\mathrm{detect}}$ for each pipeline considered, based on photometrically confirmed detections (see Appendix \ref{app:B}). The dotted curves show the model yield (as in Figure \ref{fig:cmd}) assuming a stellar mass $M=1.1\ \mathrm{M_\odot}$ not including (blue) and including (orange) dilution effects (see text), respectively. The solid black curve shows the predicted yield (including dilution) when intrinsic stellar parameters are sampled by comparing the 2MASS photometry with the PARSEC isochrones. The observed detection yield has a relatively low dependence on the number of sectors, but has a clear peak in $M_H$ and $T_{\mathrm{eff}}$, where the oscillation modes are detected best.}
    \label{fig:fdetect}
\end{figure*}

We first examine the yield of seismic detections for the bright SCVZ red giants in all three pipelines and determine the sample where all analyses provided consistent results for the global parameters. We compare realised yields with simple predictions, computed using the formalism presented in \citet{2011ApJ...732...54C} and updated for \emph{TESS} targets by \citet{2016ApJ...830..138C,2019ApJS..241...12S}, implemented for this sample as the \texttt{asteroestimate}\footnote{\url{https://github.com/jmackereth/asteroestimate}} Python package. The model uses \emph{Gaia} and 2MASS photometry, the \emph{Gaia} parallax and time series length as input, but requires an initial estimate (or prior) on stellar mass. Assuming the simple scaling for the oscillation amplitudes as in \citet{2011ApJ...732...54C}, mass has a limited impact on the  detectability of the oscillations. More complex relationships have now been demonstrated in the literature, \citep[e.g.][]{2012A&A...543A.120S,2013MNRAS.430.2313C,2018ApJS..236...42Y}, however since the uncertainty on the predicted $\nu_{\mathrm{max}}$ is likely already large, we adopt the simpler relationship. We use the PARSEC stellar evolution models \citep{2012MNRAS.427..127B,2017ApJ...835...77M} to determine a prior on the stellar mass by comparison with the 2MASS photometry of each star, as well as computing the yield assuming a fixed mass $M = 1.1\ \mathrm{M_\odot}$. Since these give similar results, we focus below on the statistics based on the mass prior from PARSEC. Importantly, we also model the effects of dilution or `wash-out' of the asteroseismic signal of the target star by the oscillations of other stars in the photometric aperture following \citet{2016ApJ...830..138C} (This may alsobe referred to as `crowding'). We implement this by finding the ratio of flux of all stars inside each target aperture in \emph{Gaia} DR2 (down to $G = 17$) to that of the target star. This ratio, $D$, is then factored into the expressions for total mean mode power and granulation power when estimating the probability of detection \citep[ see equations 5 and 12 in][]{2016ApJ...830..138C}.

We define all detection yields as the fraction of stars out of the 8,249 with \emph{Gaia}-based luminosity estimates which had successful detections of $\nu_{\mathrm{max}}$ \emph{and} $\Delta\nu$ that were confirmed using photometric luminosities. The fiducial model, accounting for the time-series length and dilution effects predicts an average detection yield of $\sim 46\%$. The overall detection yield over the seismic pipelines is $\sim 36\%$ ($\sim 2,890$ stars) across the whole sample. However, the mean observed yields are $\sim50\%$ for stars with the full 13 sectors of data. We find that there are 6388 stars ($41\%$ of the entire 15,405 star parent sample, and 1693 of which are in the `true' CVZ) that had detections in common, but not necessarily consistent, between all three pipelines. Below, we define a  `gold' sample from the subset of these stars, for which the global parameters were highly consistent. 

\autoref{fig:cmd} demonstrates the colour-magnitude diagram (CMD) of the \emph{TESS}-SCVZ bright giant sample. The coloured contours demonstrate the regions that contain approximately 90\% of the stars with detections in each pipeline. The black dashed line shows the region containing 90\% of the stars that had a high detection probability in our fiducial detection model with dilution. Each pipeline covers a region of the CMD that contains a number of interesting features, such as the red clump, giant branch and asymptotic giant branch bump (these are discussed more specifically in \autoref{sec:results}).

\begin{figure*}
    \centering
    \includegraphics[width=0.9\textwidth]{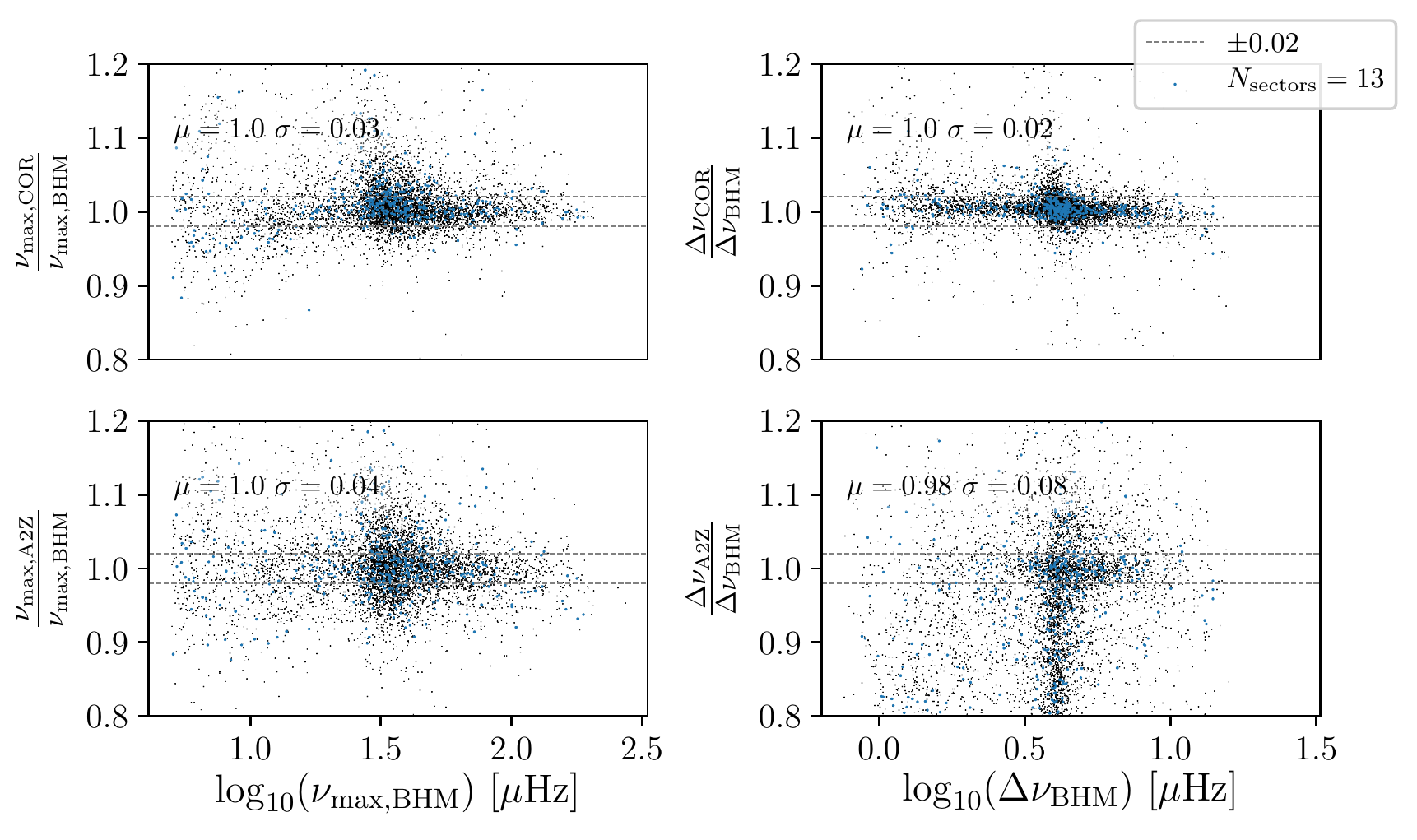}
    \caption{Internal consistency between pipelines. Left panels compare the $\nu_{\mathrm{max}}$ returned by the COR and A2Z pipelines to the measurement from the BHM pipeline. Right panels show the same for $\Delta\nu$. Only stars for which results were returned in both relevant pipelines are shown. We highlight stars that have the full 13 sectors of data from \emph{TESS}. The mean and standard deviation of the marginalised distributions are shown in the top left of each panel. In general, any differences are on the few percent level.}
    \label{fig:pipes}
\end{figure*}

\autoref{fig:fdetect} shows the yield $f_{\mathrm{detect}}$ for each pipeline (coloured histograms) as a function of number of sectors observed $N_\mathrm{sectors}$, absolute $H$-band magnitude $M_H$ and effective temperature $T_{\mathrm{eff}}$ (from SkyMapper). The coloured dashed curves show the expected yield from the detection models with and without dilution. The solid black curve shows the predicted yield for the fiducial model based on the PARSEC mass prior including dilution. While the details and performance of the individual pipelines are discussed in their respective papers, it is worth noting that they generally agree in terms of where the greatest yields can be achieved. It is clear that there is a range in $M_H$ and $T_{\mathrm{eff}}$ where the global seismic parameters can be readily measured, between $-4 \lesssim M_H \lesssim 0$ and $4000 \lesssim T_{\mathrm{eff}} \lesssim 5000$, consistent with the expected range from the model. Not allowing for dilution leads to an additional $\sim 20\%$ predicted yield in all cases, suggesting that the large pixel size of \emph{TESS} negatively affects seismic yields, as expected. Evidently, from Figures \ref{fig:cmd} and \ref{fig:fdetect}, the pipelines perform best for less evolved giants, which are fainter and hotter. These stars were more abundant in the \emph{Kepler} sample, where these pipelines have been well-tested. The effect of the high-pass filtering applied to the light curves likely also biases our analysis against the brighter giants which are predicted to have detectable oscillations. 

Small differences due to population effects are evident between the PARSEC prior and the fixed mass yield models. Higher mass targets are slightly under-detected relative to lower mass stars (this will change depending on the adopted scaling relation for $A_{\mathrm{max}}$). Proper modelling of these effects will be of particular importance, for example, when trying to ascertain the Galactic star formation history based on asteroseismic samples from \emph{TESS}, as it will impose a bias on the derived seismic age distribution against younger (and therefore more massive) targets. However, the simpler target selection of this sample makes this inherently possible, as the simple selection in colour-magnitude space is invertible using stellar population models.  

We make a brief check of the internal consistency between the measurements of the global seismic parameters of the pipelines considered here in \autoref{fig:pipes}. We compare the results from the COR and A2Z pipelines to those from BHM to assess which, if any, pipelines are internally consistent. The summary statistics are shown in the upper left of each panel. The left panels reveal that in general the pipelines agree on $\nu_{\mathrm{max}}$ at a level of 3 to 4 percent, with no significant offsets from the global mean. The average consistency in $\Delta\nu$ is smaller between COR and BHM, at a level of $\sim 2$ percent, but is hampered between A2Z and BHM due to a set of Red Clump giants whose $\Delta\nu$ measures are significantly different. We find similar inconsistencies between COR and A2Z. There is, however, still a core set of stars for which this parameter is consistent at the $<2$ percent level (indicated by the dashed grey lines in each panel). It is conceivable that $\Delta\nu$ measures should suffer for shorter time series data. However, we find that the stars with $N_{\mathrm{sectors}} < 3$ are consistent at a similar level to those with $N_{\mathrm{sectors}} = 13$. This likely indicates some issue in definition of $\Delta\nu$ between pipelines, or some systematic issue in our \emph{TESS} light curves that is affecting this measurement in the A2Z pipeline. For example, with few detectable orders, the definition of $\Delta\nu$ as the mean or median frequency spacing or some other weighting of these separations becomes important. Pipelines tend to differ in this regard and so may return significantly different results. The effect of this is clearly seen as larger inconsistency at low $\langle \Delta\nu \rangle$, and indeed in $\nu_{\mathrm{max}}$, in \autoref{fig:pipes}. Extending the time series with data from the extended \emph{TESS} mission will improve the data in this regard, but there is almost certainly fine-tuning that still needs to be applied to achieve the best results from all pipelines.

To determine a sample of stars where all pipelines return consistent parameters, we compute the global mean of parameter measurements in cases where parameters were returned from all three pipelines. There are 5,574 stars (1521 in the `true' CVZ) for which the parameter measurements from all pipelines were consistent with the global mean within their combined uncertainty. This subset is somewhat reduced from the 6,388 stars with detections in all pipelines. Much of this is driven by inconsistencies in $\Delta\nu$ noted above. The relatively large number of stars entering this `gold' sample indicates that the pipelines are estimating uncertainties well. As an example, the median combined uncertainty on $\nu_{\mathrm{max}}$ is $1.73\ \mathrm{\mu Hz}$ or $\sim 5$ percent. For $\Delta\nu$ the median uncertainty is $0.12\ \mathrm{\mu Hz}$, or $\sim 3$ percent. Determinations of $\Delta\nu$ from individual pipelines generally have lower uncertainties than the mean values. For the remainder of the paper, we choose to use the BHM values as standard in order to demonstrate the prospects for \emph{TESS}, since these were the results with the greatest consistency with the COR pipeline. {\red The selection of a single pipeline does not significantly affect the results of the following sections, but BHM makes a natural choice since its measured $\Delta\nu$ is defined similarly to that derived for the stellar models \citep{Elsworth_2020}. Furthermore, in testing, we find that there is little difference (less than $\sim 0.3\sigma$) between the $\Delta\nu$ values returned by BHM and those catalogued by \citet{2018ApJS..236...42Y} for stars in the \emph{Kepler} field. The \citet{2018ApJS..236...42Y} results agree very closely with those from individual mode frequencies and so are a good benchmark \citep{2019A&A...628A..35K}.}  Despite this selection, we still include the global mean values and the seismic gold sample identifier in our published catalogue (see Table \ref{tab:datamodel} in Appendix \ref{app:D}), recommending these stars as a benchmark for the \emph{TESS}-SCVZ, but noting that these early values should be used with certain cautionary steps, also outlined briefly in Appendix \ref{app:D}.

\subsubsection{Galaxia model}

We assess the future potential for seismic samples using \emph{TESS} data with a simple model of the SCVZ generated using \texttt{Galaxia} \citep{2011ApJ...730....3S}. \texttt{Galaxia} generates realistic stellar populations with realistic spatial distributions by sampling from a density model fit to Milky Way data and stellar models. For the model, we compute the 2MASS and \emph{Gaia} photometry using \texttt{bolometric-corrections}. We then compute the detection yield in the model using the procedure above (using the stellar parameters given by \texttt{Galaxia}), extending the sample down to fainter magnitudes to gain an insight into the statistics available for fainter $G$ magnitude stars.  Again, we include dilution by other sources within the target apertures. We use the parameters provided by the stellar models in \texttt{Galaxia} \citep[which are derived from the PARSEC library: ][]{2012MNRAS.427..127B,2017ApJ...835...77M,2017MNRAS.467.1433R} to estimate $\nu_{\mathrm{max}}$ for each star in the model via the usual asteroseismic scaling relations and use the mass provided as input to \texttt{asteroestimate}.

\begin{figure}
    \centering
    \includegraphics[width=0.9\columnwidth]{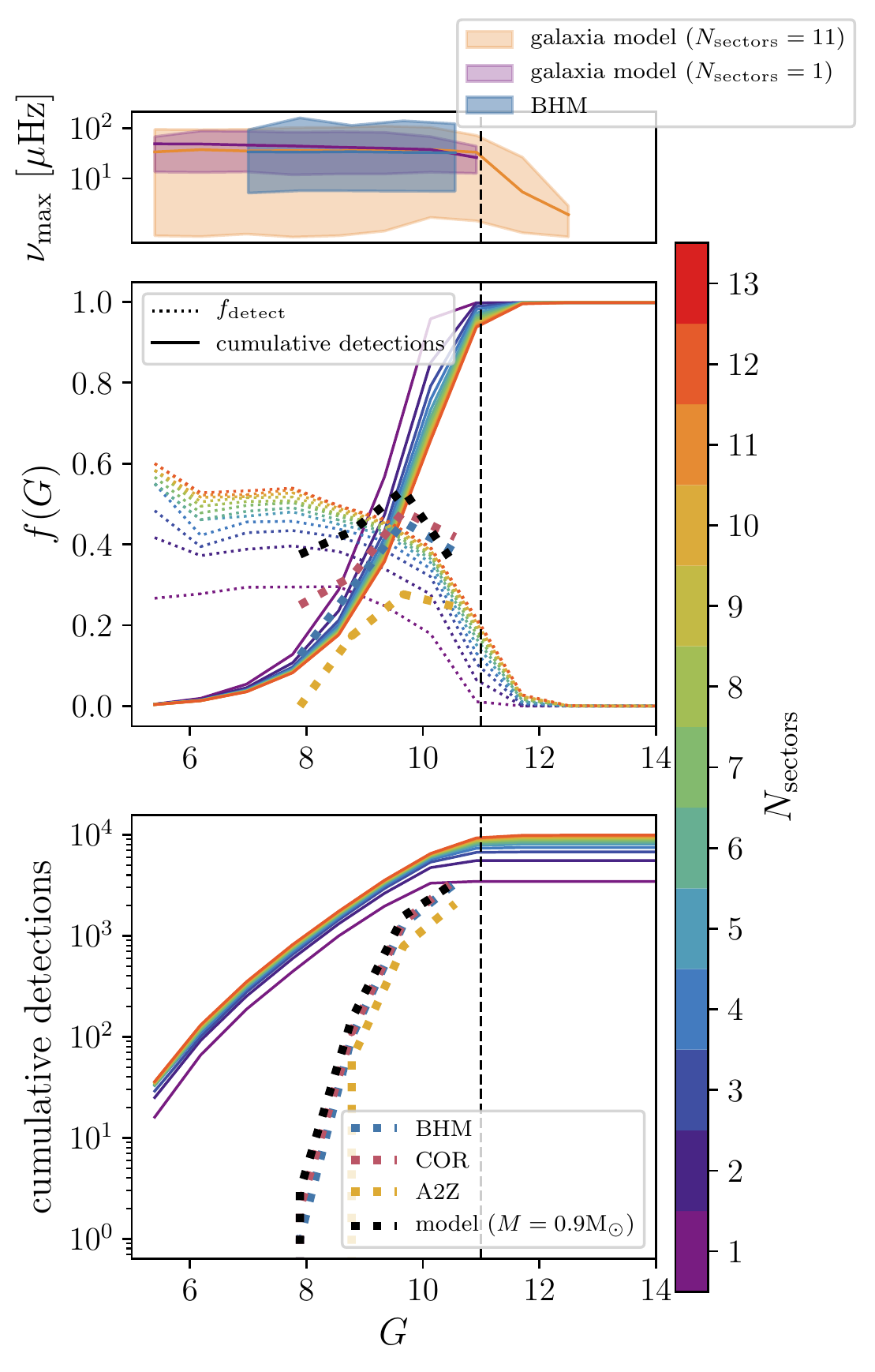}
    \caption{Detection statistics from a \texttt{Galaxia} model of \emph{TESS}-SCVZ giant stars selected using the same photometric cuts as our giants, compared to the realised detection yields (and predictions) of the observed bright giants. The top panel shows the predicted detectable and observed (by the BHM pipeline) range of $\nu_{\mathrm{max}}$ as a function of $G$ and for $N_{\mathrm{sectors}}=1$ and 11. The middle panel shows the detection yield $f_\mathrm{detect}$  (dashed lines) and cumulative fractional detections (solid lines) as a function of $N_\mathrm{sectors}$. The lower panel gives the absolute cumulative number of detections as a function of $G$. We compare the model yields with the observed and predicted yields based on our observational sample, which are shown by the wide dashed lines (coloured as in \autoref{fig:cmd} and \ref{fig:fdetect}) in the lower two panels. }
    \label{fig:galaxia}
\end{figure}

\autoref{fig:galaxia} summarises the statistics from our \texttt{Galaxia} model in comparison with our realised and modelled yields. The top panel demonstrates the range of $\nu_{\mathrm{max}}$ for which seismic detections are theoretically possible as a function of $G$, compared to those detected by the BHM pipeline (we note that this range is nearly identical for all pipelines). Our model predicts detections down to well below $\nu_{\mathrm{max}} < 10\ \mathrm{\mu Hz}$ for the $N_{\mathrm{sectors}} > 11$ data. Extending the sample to fainter magnitudes than the currently adopted $G = 11$ limit (demonstrated by the vertical dashed line) will require better characterisation of these intrinsically low $\nu_{\mathrm{max}}$ targets. At fainter magnitudes, the observable giants are dominated by intrinsically bright (and therefore low $\nu_{\mathrm{max}}$) upper RGB and AGB stars. 
For stars with $N_{\mathrm{sectors}} = 1$ the predicted detectable range of $\nu_{\mathrm{max}}$ is significantly decreased, owing to the fact that low frequency oscillations are not well sampled by the shorter time series. 

The lower two panels of \autoref{fig:galaxia} demonstrate the attainable detection yields (middle panel) and absolute sample size (lower panel) predicted by the \texttt{Galaxia} model in our adopted magnitude range, and at fainter magnitudes. The middle panel compares the predicted yield from \texttt{Galaxia} for different $N_\mathrm{sectors}$ (fine dashed lines) with those realised from our pipelines and predicted by the model based on observations. The solid curves show the cumulative fraction of detections as a function of $G$, indicating the fraction of all possible detections made at each $G$ limit. At faint $G\gtrsim 9$, the realised yields agree well with those from \texttt{Galaxia}, suggesting that yield predictions based on the \texttt{Galaxia} model at fainter magnitudes are trustworthy. The yield turns over strongly at $G > 11$, suggesting this is roughly the limit for detectable giants in \emph{TESS}. The \texttt{Galaxia} model suggests that there are many detectable brighter giants, however. The absolute cumulative detection counts shown in the lower panel also reflect this. It is likely that least some of these bright stars may be missed in our parent sample due to selection effects in \emph{Gaia} \citep{2020arXiv200508983B} and our own imposed limits on parallax SNR, for example. Analyses of bright targets in \emph{Kepler} have shown that such stars require specialised reduction \citep{2017MNRAS.471.2882W}, and this is likely also true for \emph{TESS}.

This simple model demonstrates that somewhat increased seismic sample sizes are attainable at fainter magnitude limits, albeit at the cost to the detection `hit-rate'. Based on our observed yields, we predict that there are likely a few hundred more seismically detectable giants in the SCVZ, particularly at brighter $G$. Assuming the NCVZ has similar statistics, across the north and south the potential seismic CVZ giant sample could be as large as $\sim 10^4$ stars. Based on this model, we predict that seismic detections can be made for $\sim 3\times10^5$ giants across the whole sky (i.e. for $N_{\mathrm{sectors}} = 1$, subject to the limits on $\nu_{\mathrm{max}}$ outlined above \citep[in good agreement with the expectations of][]{2019arXiv191207604S}.

\subsection{Constraints on stellar mass and age}

Using \texttt{PARAM}, we make constraints on mass and age for the set of stars with BHM results that had consistent $\nu_{\mathrm{max}}$ and $\Delta\nu$ with the COR and A2Z pipelines, and that had seismic luminosities consistent with those derived from \emph{Gaia} (i.e. \texttt{numax\_dnu\_consistent} $=1$ and \texttt{lum\_flag\_BHM} $=1$ when using the catalogue presented in Appendix \ref{app:D}). The resulting sample has 1,749 stars, although we include \texttt{PARAM} results in the final catalogue for every target that had BHM results and \emph{SkyMapper}/\emph{TESS}-HERMES parameters which did not meet these criteria (4,453 stars).  Additionally, we provide \texttt{PARAM} results for the 351 of the 513 stars with APOGEE spectra that had robust BHM results, using the APOGEE catalogue stellar parameters as constraints. 

\begin{figure}
    \centering
    \includegraphics[width=0.8\columnwidth]{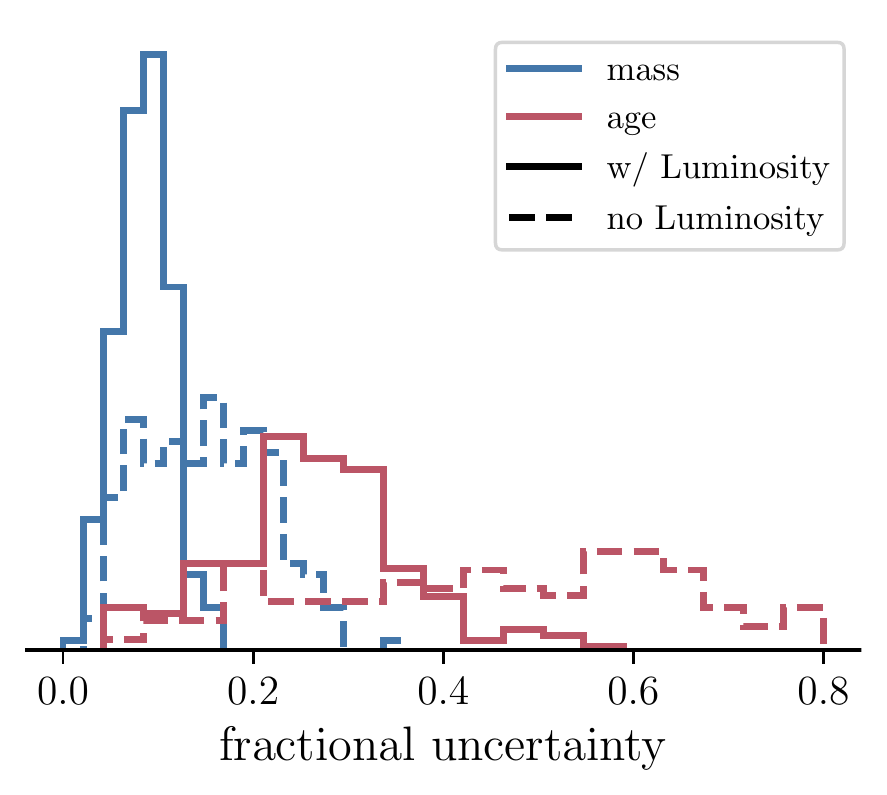}
    \caption{Histograms of fractional formal uncertainties on mass (blue) and age (red) from \texttt{PARAM} using \emph{SkyMapper}/\emph{TESS}-HERMES stellar parameters and BHM seismic results. We compare the effect of including \emph{Gaia} luminosity as a constraint between the dashed (no luminosity) and solid (including luminosity) histograms for age and mass.  Inclusion of the \emph{Gaia} luminosity improves the median uncertainties from 15\% to 8\% in mass, and  50\% to 26\% in age.}
    \label{fig:massagelum}
\end{figure}

Since the seismology results are likely to not yet be optimal, we test how the inclusion of \emph{Gaia} luminosity, $L_{\mathrm{phot.}}$, as a constraint for \texttt{PARAM} influences the accuracy and precision of our mass and age constraints. The median precision on $\nu_{\mathrm{max}}$ and $\Delta\nu$, at a level of 5\% and 3\%, will not currently provide adequate constraints on mass and age without additional constraints, although individual pipelines report measurements of these parameters at much higher precision. In \autoref{fig:massagelum} we show the distribution of fractional uncertainty in mass and age for the 1,749 star sample described above.  These uncertainties are `internal' in the sense that they assume that the grid of stellar models used in \texttt{PARAM} are correct in an absolute sense. Dashed histograms show uncertainties when  $L_{\mathrm{phot.}}$ is not included as a constraint in \texttt{PARAM}, whereas the solid histograms show the same when  $L_{\mathrm{phot.}}$ is included. The uncertainties are clearly tightened up when luminosity constraints are included. Formally, the median uncertainty on mass decreases from 15\% to 8\%, and on age from 50\% to 26\%. Without using  $L_{\mathrm{phot.}}$, the age uncertainties have a large spread, with many stars exceeding 70\% uncertainties, whereas its inclusion reduces 93\% of the stars to uncertainties of less than 40\%. Clearly, improving precision and accuracy on luminosities with better spectroscopic data, for example, will improve our constraints on mass and age. Furthermore, this demonstrates that the \emph{TESS} seismic results are not yet optimal, and can be improved via refinements of the light curves and analyses of them.

For the remainder of the paper, we focus solely on the results which use $L_{\mathrm{phot.}}$ as an additional constraint. Furthermore, we include only these results in our final catalogue, as they are likely to be the most robust. We refer to results based on the \emph{SkyMapper}/\emph{TESS}-HERMES parameters as $\mathrm{age_{SM}}$ and $M_{\mathrm{SM}}$, and those based on APOGEE spectroscopic constraints as $\mathrm{age_{APO}}$ and $M_{\mathrm{APO}}$.

\subsection{Stellar Astrophysics}

\begin{figure*}
    \centering
    \includegraphics[width=0.9\textwidth]{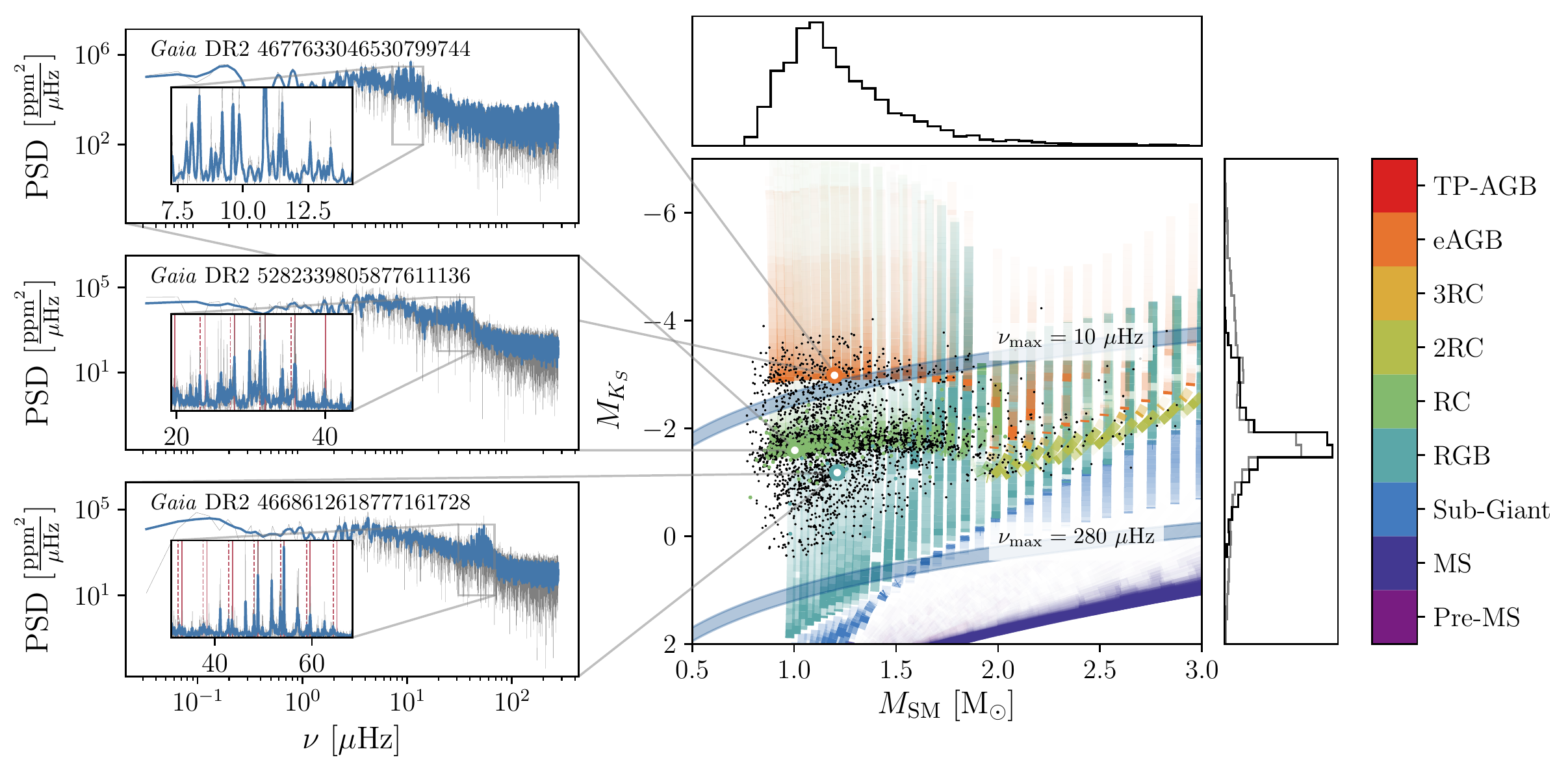}
    \caption{\emph{Left:} Examples of the seismic detections made by \emph{TESS} across the evolutionary stages of giants in the CVZ. Clear power excess is seen in the log-log space spectra, and oscillation modes are resolved within these envelopes. The $l=0$ and $l=2$ modes are shown by the red solid and dashed vertical lines (respectively) in the lower two panels. The top panel shows a candidate AGB bump star, which has a complicated mode structure. \emph{Right:} Mass against absolute $K_S$ band magnitude $M_{K_S}$ for CVZ giants with detections in the BHM pipeline (which have a seismic luminosity consistent with that from \emph{Gaia}), compared with a set of solar metallicity isochrones from PARSEC, with the evolutionary stages demonstrated in colour. The Red Clump phase is divided into three stages RC, 2RC, 3RC, which represent the core Helium burning phase of low, intermediate and high mass stars, respectively. Red Clump giants whose state is confirmed using the methods of \citet{2016A&A...588A..87V} are shown as green scatter points. The histograms demonstrate the marginalised distribution of mass and $M_{K_S}$ (the full underlying distribution of $M_{K_S}$ is shown in gray). The positions of the example stars in this space are indicated by large white points, and the $M_{K_S}$ range at which $\nu_{\mathrm{max}}$ is equal to 10 and 280 $\mathrm{\mu Hz}$ at each mass (for a representative range in $T_{\mathrm{eff}}$) is shown by the labeled blue bands. }
    \label{fig:mass_mk}
\end{figure*}

The potential for \emph{TESS} to provide accurate asteroseismic parameters for many giants on an all-sky basis has been shown for single-sector 27-day data by \citet{2019arXiv191207604S}. The year-long data provided by targets in the CVZ in the North and South allows detections of asteroseismic signal at lower frequencies, making the seismic parameters of brighter (and thus larger radii) giants more readily measurable. Longer time-series also afford higher frequency resolution, which allow features in the spectra from mixed modes and rotational splitting to be measured \citep[e.g.][]{2019A&A...622A..76M}. Shorter time series make such analyses more difficult \citep[e.g.][]{2019ApJ...885...31C}.
In this section, we demonstrate the prospects for new constraints on models of stellar structure and evolution using asteroseismology of giants in the 13 sectors of data in the CVZ.

\autoref{fig:mass_mk} shows examples of power spectra for stars in the \emph{TESS}-SCVZ sample at different evolutionary stages. The right panel shows stellar mass $M$ as estimated using the BHM seismic and SkyMapper/\emph{TESS}-HERMES spectroscopic parameters in \texttt{PARAM} against the absolute $K_S$-band magnitude $M_{K_S}$, computed using the \emph{Gaia} DR2 parallaxes.  The lines plotted underneath show evolutionary tracks from the PARSEC stellar evolution models \citep{2012MNRAS.427..127B,2017ApJ...835...77M}. Each line is coloured by the evolutionary stage as provided in the PARSEC tracks and shaded to reflect the relative time spent by a star at each point, emphasising the phases where an overdensity of stars should be expected. The mass-$M_{K_S}$ distribution of the \emph{TESS}-SCVZ giants roughly matches the prediction of the PARSEC tracks, with the Red Clump (RC) clearly visible. There is also an overdensity at brighter $M_{K_S}$ than the RC, which corresponds with the Asymptotic Giant Branch (AGB) phase in the PARSEC tracks. Many of these stars are likely to be in the AGB 'bump'. This feature was originally highlighted in stellar tracks by \citet{1978MNRAS.184..377C}, observed in external galaxies by \citet{1998ApJ...495L..43G} and recently characterised in \emph{Kepler} by \citet{2015MNRAS.453.2290B}.

The left hand panels of \autoref{fig:mass_mk} show the power spectra of the stars indicated in the right panel, which were selected as examples of the main evolutionary stages which are represented in the data. The top panel shows a star with a very low $\nu_{\mathrm{max}}$ ( $< 12\ \mathrm{\mu Hz}$), at an $M_{K_S}$ and $M_{G}$ value consistent with that of AGBb stars . Oscillation modes are clearly present, but have a more complex structure \citep[since they are further from the asymptotic regime, e.g.][]{2014ApJ...788L..10S}. This complexity is evident in modelled oscillations of such stars. 
The luminosity ratio between the RC and the AGBb is of importance in constraining the size of the C-O core at the end of the core-He burning phase, given its weak dependence on the metallicity and initial helium abundance \citep[e.g.][]{1995A&A...297..115B}. The \emph{TESS}-CVZ bright giant sample provides a largely unbiased sample of these early AGB stars for which detailed seismic inference is possible. The target selection of e.g. \emph{Kepler} was complex and biased against luminous stars \citep[see, e.g.][]{2014ApJS..215...19P,2015ASSP...39..125G,2016ApJ...822...15S}, while K2 only measures short time-series and therefore does not provide the necessary frequency resolution \citep[see, e.g. Figure 1 of][]{2015ApJ...809L...3S}.

The lower two panels on the left hand side of \autoref{fig:mass_mk} show spectra of exemplar RC and Red Giant Branch (RGB) stars. Again here, oscillation modes are clearly detected in both stars. RGB and RC stars have been examined in detail in previous asteroseismic samples from e.g. CoRoT, \emph{Kepler} and also in \emph{TESS}. For this reason, our prior knowledge of their mode structure is good, facilitating their automated detection (peak-bagging) using the \texttt{PBJam}\footnote{\url{https://github.com/grd349/PBjam}} code \citep[][submitted]{nielsen2020}. The $\ell=0$ and $\ell=2$ modes are cleanly detected and are shown by the red vertical lines in the inset panels. $\ell=1$ modes are visible in between the $l=0,2$ pairs, with clear evidence of mixed pressure and gravity modes visible within each order. While we do not make use of individual mode measurements in our analysis, this demonstrates that such analyses are inherently possible with \emph{TESS}. 

\begin{figure*}
    \centering
    \includegraphics[width=0.85\textwidth]{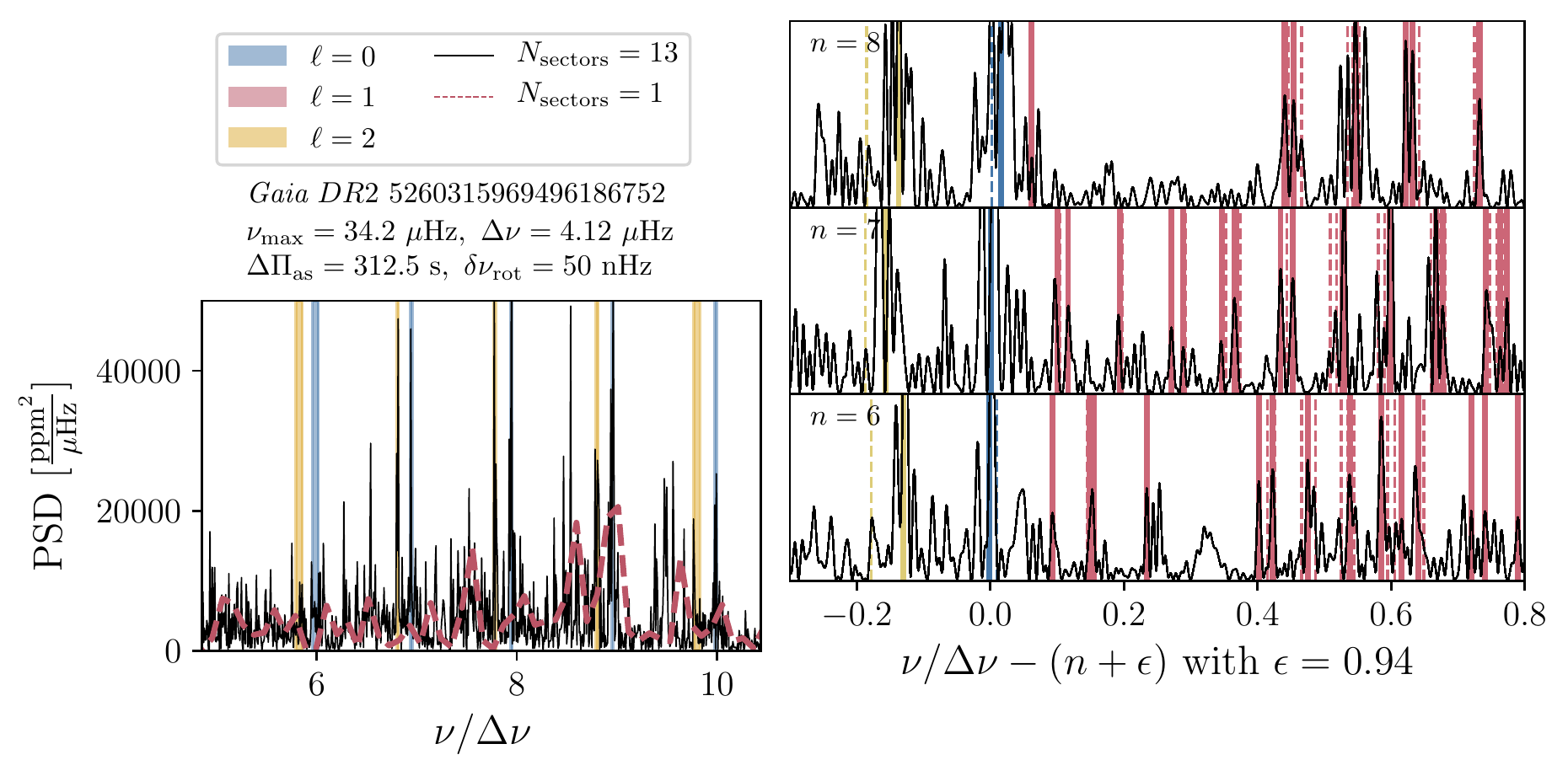}
    \caption{Example of mixed modes visible in a red clump giant identified using the period spacing method of \citet{2016A&A...588A..87V}. The spectrum around the power envelope is shown on the left, with zoomed sections at each radial order on the right. The $\ell=0,2$ modes (identified with \texttt{PBJam}) are indicated by the blue and yellow bands. At each order, a `forest' of $\ell = 1$ modes is clearly visible, split into a number of separated peaks. The power spectrum of the same star based on 1 sector of data is shown by the red dashed line, demonstrating that these modes would not be detected in that data. A fit to the oscillation modes following the methods of \citet{2015A&A...584A..50M,2018A&A...618A.109M} is shown by the dashed and solid vertical lines in the right panel. Solid lines show the observed modes, whereas dashed shows the prediction of the asymptotic relation. Blue and yellow bands indicate  $\ell=0,2$ modes, as before. Red lines show the positions of the $\ell=1$ mixed modes detected in the fit, which are split by stellar rotation at a level of $\delta\nu_{\mathrm{rot}} = 50\ \mathrm{nHz}$.  }
    \label{fig:modesplit}
\end{figure*}

We further demonstrate the ability of the \emph{TESS}-CVZ data to study mixed modes and rotational splitting in red giant stars in \autoref{fig:modesplit}. We show the power spectrum around $\nu_{\mathrm{max}}$ for an RC giant identified using methods outlined in \citet{2016A&A...588A..87V}. In the left hand panel, the nearest five radial orders to $\nu_{\mathrm{max}}$ are shown. The right panel shows the modes at $n=6,7,8$, and the results of a fit to the $\ell=0,1$ and $2$ modes following \citet{2015A&A...584A..50M,2018A&A...618A.109M}. The solid black line shows the power spectrum based on the full 13 sector data set, whereas the underlying dashed red line shows the same generated based on just 27 days (a single sector) of data. In the left panel, we identify the positions of $\ell=0,2$ modes as fit by \texttt{PBjam} (blue and yellow vertical bands, width indicates 95\% confidence interval). The structure between these modes is due to $\ell = 1$ dipole modes. It is evident in the 13 sector curve that these modes divide into multiple separated peaks, whereas such behaviour is not visible in the single sector data (dashed line). In red giant stars, mixed modes are generated by coupling of pressure and gravity modes, which is related to the density contrast between the core and the convective envelope of the star  \citep[e.g.][]{2010ApJ...721L.182M, 2011Natur.471..608B,2011Sci...332..205B,2012A&A...540A.143M}. The fit to the dipole modes in the right panel shows that not only do we detect mixed modes, but also tentatively detect splitting within these modes due to stellar rotation. The best fit mean rotational splitting of the mixed modes is $\delta\nu_{\mathrm{rot}} = 50\ \mathrm{nHz}$, with a similar splitting indicated across multiple orders. It is important to note that this value is just above the threshold of $\delta\nu_{\mathrm{rot}} \simeq 30\ \mathrm{nHz}$ implied by the frequency resolution.  Our detection of rotational splitting in an RC star here is comparable in scale to initial measurements of red giants based on $\sim 500$ days of \emph{Kepler} data \citep{2012Natur.481...55B}.

\begin{figure}
    \centering
    \includegraphics[width=0.9\columnwidth]{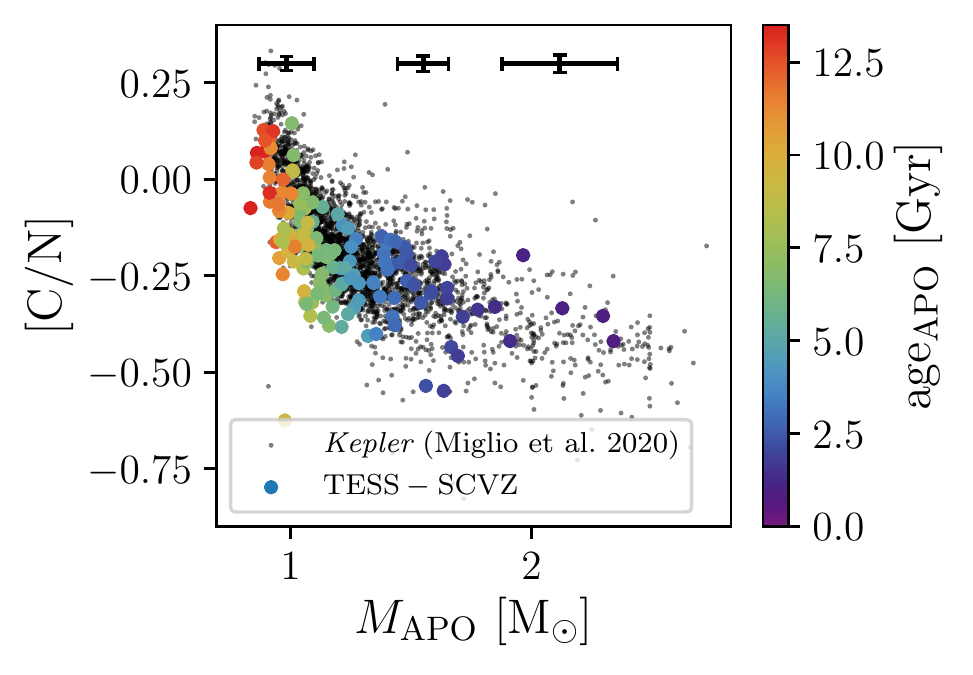}
    \caption{Mass vs the ratio of Carbon and Nitrogen abundances, $\mathrm{[C/N]}$, for the 351 \emph{TESS}-CVZ stars with results from \texttt{PARAM} based on APOGEE spectroscopic constraints and BHM seismic results, compared to the same relationship from the study of \emph{Kepler} stars by \citet{2020arXiv200414806M} (shown as small black points). The \emph{TESS}-SCVZ points are coloured by the age derived from \texttt{PARAM}, and the median uncertainty on Mass and $\mathrm{[C/N]}$ is demonstrated at three characteristic masses by the black error bars.  The \emph{TESS}-SCVZ results broadly follow the same trends as those from \emph{Kepler}.}
    \label{fig:masscn}
\end{figure}

Finally, in \autoref{fig:masscn},  we demonstrate that the trend between stellar mass and ratio of the photospheric Carbon and Nitrogen abundances, $\mathrm{[C/N]}$, which is readily seen in APOKASC and other data-sets \citep[e.g.][]{2016MNRAS.456.3655M,2015MNRAS.453.1855M,2019arXiv191207604S}, is also broadly reproduced by our \emph{TESS}-SCVZ results. Here, we use the subset of stars in our sample which have APOGEE spectra, that were re-analysed in PARAM with these tighter spectroscopic constraints. We compare with the dataset of \citep{2020arXiv200414806M}, which used a very similar analysis to derive stellar mass. The $\mathrm{[C/N]}$ has a relationship with mass in giant stars due to the first dredge-up (FDU) of material from the stellar interior as the star evolves away from the main sequence \citep[e.g.][]{2015A&A...583A..87S,2017A&A...601A..27L}. Burning of Hydrogen by the CN and then CNO cycle while on the main sequence produces Nitrogen in the stellar core. The convective envelope during the FDU reaches deeper into the interior for more massive stars, which increases the surface $\mathrm{[C/N]}$ on the RGB. This effect is clearly seen in the \emph{TESS}-CVZ data.

\subsection{Galactic Archaeology}

Stellar ages are essential, alongside element abundance information and kinematics, to the understanding of the formation and evolution of our Galaxy. As we have shown, \emph{TESS} can provide reliable seismic constraints for $\sim 10^{5}$ giants in the nearby Milky Way, across the whole sky, and $\sim 10^{4}$ with at least 1 year of data. In combination with spectroscopic constraints and modelling, seismic parameters derived from these light curves will provide these essential age constraints. Seismology and stellar modelling presently provides one of the most precise means by which to determine stellar ages \citep[with success already in \emph{TESS} e.g.][]{2019arXiv191207604S}, and perform detailed studies of the formation and evolution of the Galaxy \citep[e.g.][]{2013MNRAS.429..423M,2016MNRAS.455..987C,2016arXiv160407771A,2016arXiv160804951A,2018MNRAS.475.5487S,2020arXiv200414806M,2020arXiv200601783M}. These accurate age constraints then provide ideal training sets for data-driven methods of age estimation which can be applied to even larger data sets \citep[e.g.][]{2019MNRAS.489..176M,2019MNRAS.484..294D,2019ApJ...871..181H,2020arXiv200303316C}. Here we present a prospective look at the Galactic Archaeology potential of \emph{TESS}.

\begin{figure}
    \centering
    \includegraphics[width=1.\columnwidth]{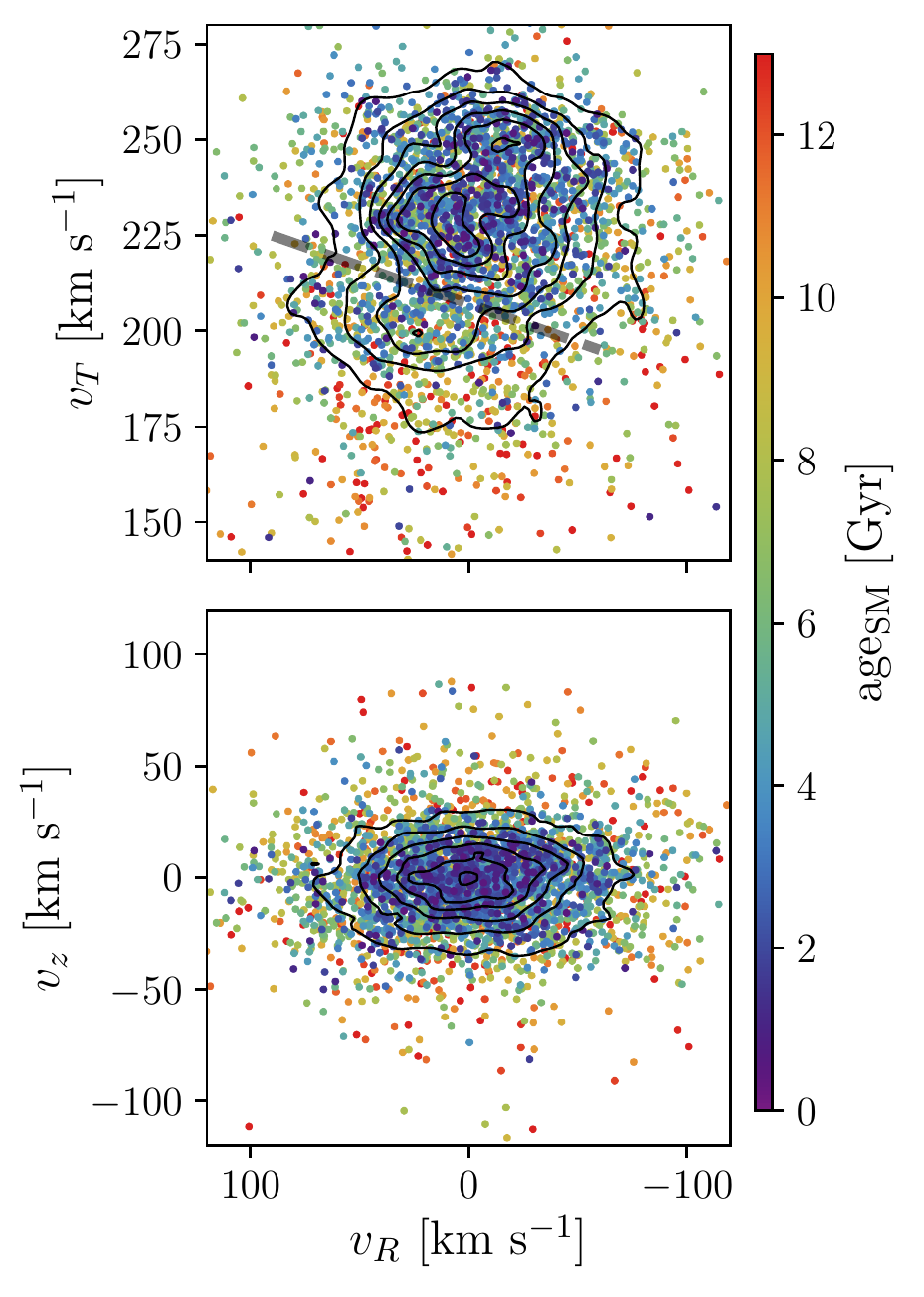}
    \caption{Kinematics of the \emph{TESS}-SCVZ sample with age estimates based on the BHM seismic constraints and \emph{SkyMapper} stellar parameters. The top panel shows the $v_{R}-v_{T}$ plane, and the bottom shows $v_{R}-v_{z}$. Points show each star from the sample with reliably measured ages, coloured by the age estimate from \texttt{PARAM}, plotted in reverse order, such that the youngest stars appear in the foreground. The contours demonstrate the kinematics of the full underlying sample. The commonly discussed velocity substructures in the disc (e.g. the \emph{Hercules} stream/gap feature at $v_R < 0\ \mathrm{km\ s^{-1}}$, $v_T < 210\ \mathrm{km\ s^{-1}}$) are visible in the underlying sample, and we indicate this in the top panel by the thick dashed line. The stars for which we have measured ages appear to sample these features well.}
    \label{fig:kinematics}
\end{figure}

We first show the kinematics of the sample in Figure \ref{fig:kinematics}. The coloured points show the radial to tangential velocity $v_{R}-v_{T}$ (top) and radial to vertical velocity $v_{R}-v_{z}$ (bottom) distribution of the sample with robust ages, determined using the BHM seismic constraints and \emph{SkyMapper} parameters, with the colour corresponding to the median age of the posterior distribution from \texttt{PARAM}. The contours show the same kinematics for the full underlying sample of \emph{TESS}-CVZ giants. Many of the commonly discussed velocity space substructures at the solar vicinity are visible in the full sample. In particular, the \emph{Hercules} stream/gap \citep[e.g.][]{1998AJ....115.2384D} is visible in the lower left of the distribution of the top panel. A tentative inspection of stars either side of this gap reveals that those at low $v_T$ seem to have slightly older ages. Recent studies of moving groups in age space have shown that there may be important constraints on the disk dynamics to be made based on relative ages such as these \citep{2020arXiv200613876L}. Furthermore, the \emph{TESS}-CVZ sample, and larger samples enabled by the use of 1-sector \emph{TESS} data or fainter stars will provide new
constraints on the local age-velocity dispersion relation \citep[also recently studied in detail by e.g.][]{2018arXiv180803278T,2019MNRAS.489..176M,2020arXiv200414806M}. In the bottom panel of \autoref{fig:kinematics} the eldest stars clearly have the largest dispersion in $v_R$ and $v_z$.

\begin{figure*}
    \centering
    \includegraphics[width=0.7\textwidth]{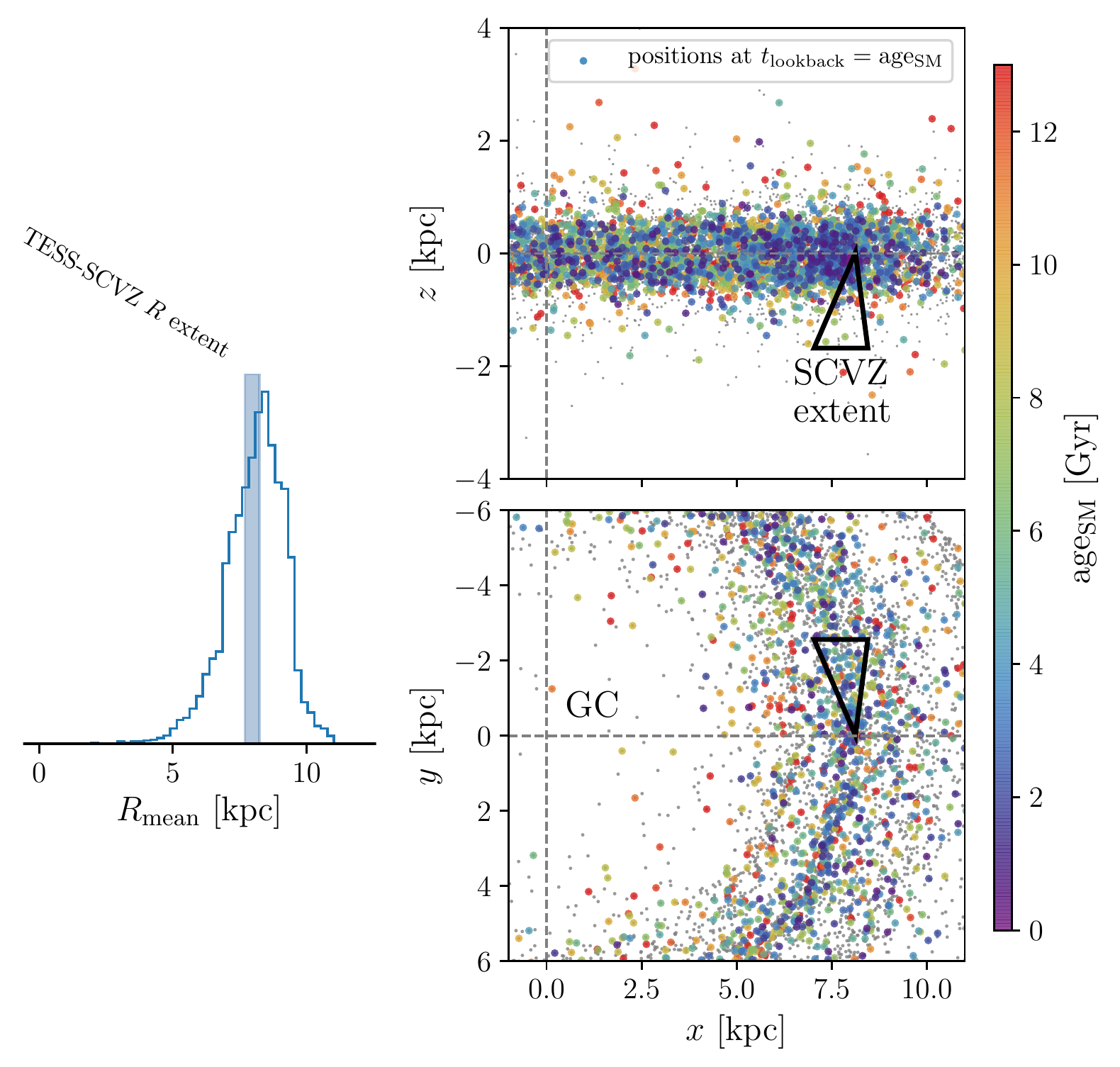}
    \caption{The spatial extent of the \emph{TESS}-SCVZ sample compared with the extent which is 'dynamically' sampled by the data. The histogram on the left compares the extent in Galactocentric $R$ today (blue shaded region) with the distribution of mean orbital radii $R_{\mathrm{mean}}$ in the sample (blue histogram). The right panels show the spatial distribution in Galactocentric Cartesian coordinates when the sample is integrated backwards to its apparent age (coloured points) or 2 Gyr for stars with no age estimates (small black points) in an axisymmetric potential. The approximate extent today of the SCVZ sample is shown by the black triangles.}
    \label{fig:spatial}
\end{figure*}

\begin{figure}
    \centering
    \includegraphics[width=0.8\columnwidth]{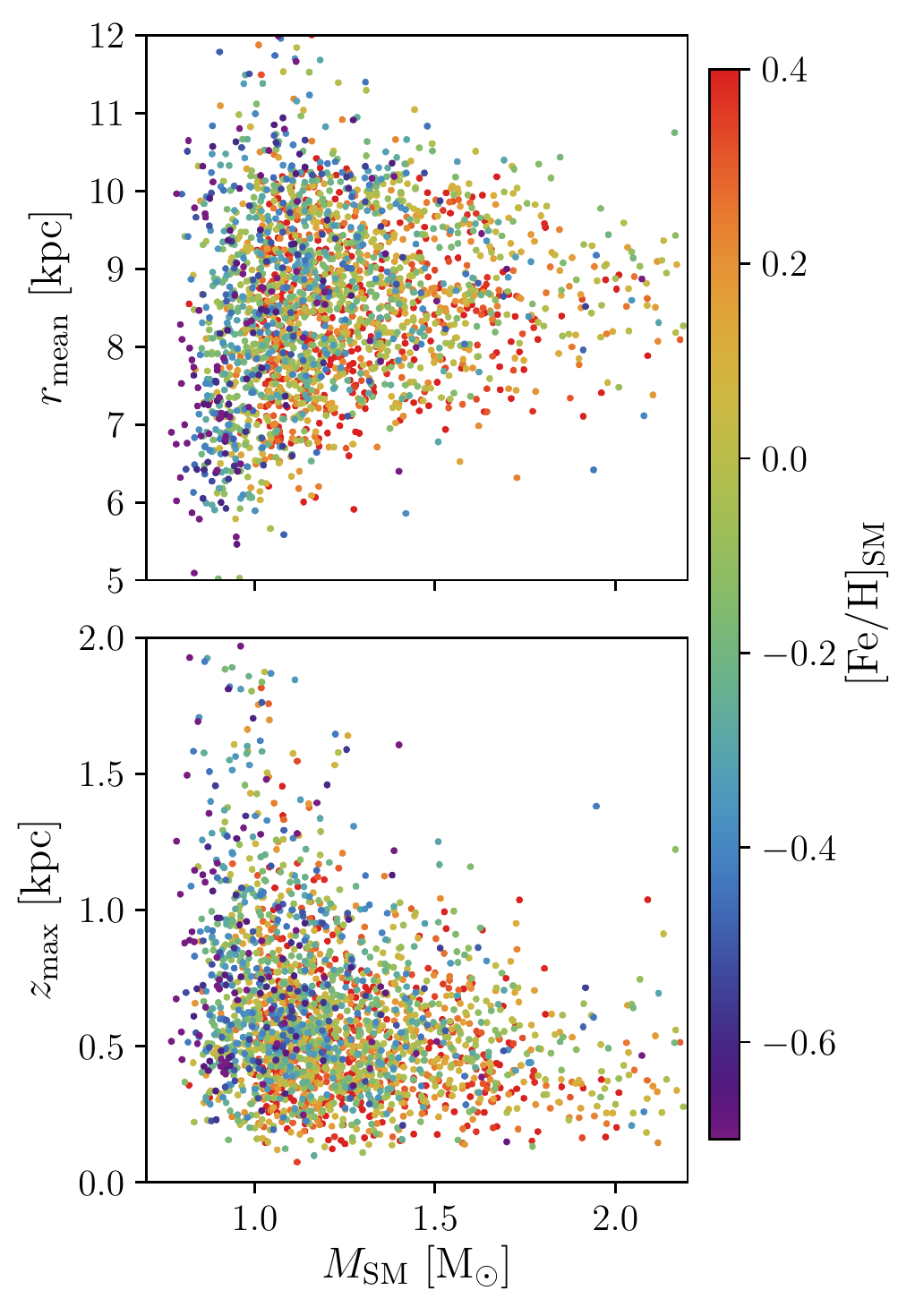}
    \caption{The stellar mass as estimated using \texttt{PARAM} applied to the BHM pipeline and \emph{SkyMapper} parameters, against the mean orbital radius $r_{\mathrm{mean}}$ and the maximum vertical excursion above the plane $z_{\mathrm{max}}$. The colour of the points is determined by the $\mathrm{[Fe/H]}$ from \emph{SkyMapper}. It is clear that the lowest mass (and therefore oldest, on average) stars are those whose orbits cover the greatest radial and vertical extents. There is a clear relationship with $\mathrm{[Fe/H]}$, such that at higher mass (and thus younger ages), stars are generally more metal rich.}
    \label{fig:zmaxrmean}
\end{figure}

In \autoref{fig:spatial}, we demonstrate how even the limited spatial extent of the \emph{TESS}-SCVZ sample still `dynamically' samples a large part of the Galactic disc. Using the kinematic information from \emph{Gaia}-DR2 shown above, we integrate the present day orbits of each star backwards in \texttt{galpy}'s \texttt{MWPotential2014} axisymmetric, static potential. For stars with age estimates from \texttt{PARAM} (using the BHM pipeline results and \emph{SkyMapper} parameters), we integrate to that age. For stars with no age estimate, we integrate to 2 Gyr ago, which is the typical age of stars for which ages could be determined. In the left panel of \autoref{fig:spatial}, we show the resulting distribution of mean orbital Galactocentric radii $R_{\mathrm{mean}}$ as compared to the present day radial extent (shown by the vertical black lines). The right panels demonstrate the spatial positions in the Galaxy (in an edge on and face on view) of the sample after the back-integration, in comparison with the approximate present day spatial extent of the sample, shown by the black triangles. It is clear that the stars present in the \emph{TESS}-SCVZ are members (in a dynamical sense) of a far more widely distributed population. While this is of course true for any local sample or field, it emphatically demonstrates the richness of the seismic data available.

\autoref{fig:zmaxrmean} shows the trends between the stellar mass (which is determined more precisely than the age in most cases) and the mean orbital radius $r_{\mathrm{mean}}$ and maximum vertical excursion from the midplane $z_{\mathrm{max}}$. At increased stellar mass (and therefore younger ages on average), the width of the $r_{\mathrm{mean}}$ distribution decreases markedly, such that the lowest mass stars have orbits whose mean radii cover the largest extent of the disc. This is a clear signal of radial mixing of disc stars, and has been noted in a number of studies \citep[e.g.][]{2011A&A...530A.138C,2018ApJ...865...96F,2019ApJ...884...99F,2020arXiv200204622F,2020arXiv200414806M,2020arXiv200503646S}. Similarly, at lower mass, stars have a wider distribution in $z_{\mathrm{max}}$, a clear example of dynamical heating of the disc \citep[e.g.][]{2016MNRAS.462.1697A,2018arXiv180803278T,2019MNRAS.489..176M}. The measurement of precise ages for nearby stellar samples provides an ideal dataset for testing models of radial mixing and dynamical heating processes. 

\begin{figure*}
    \centering
    \includegraphics[width=0.7\textwidth]{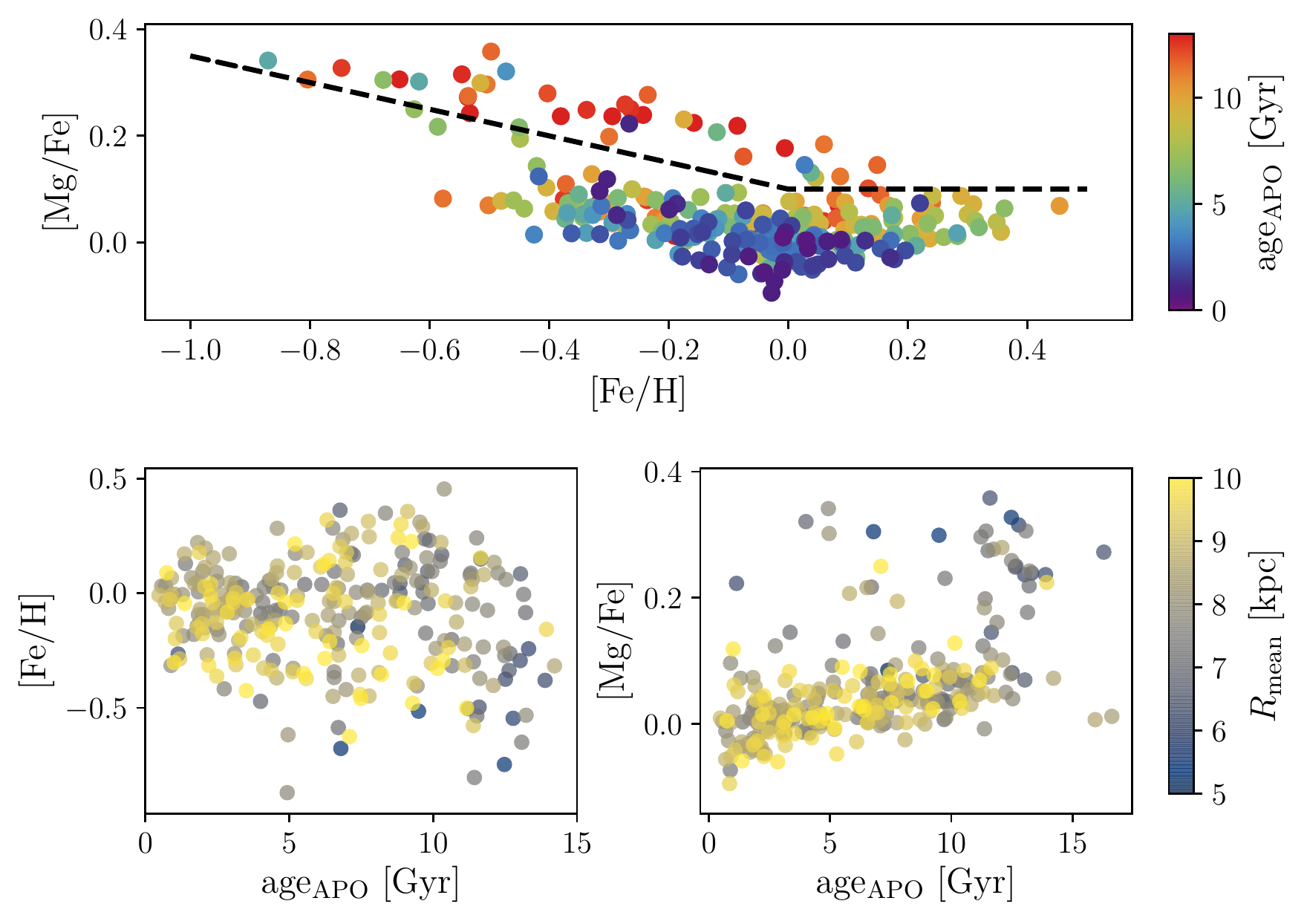}
    \caption{$\mathrm{[Mg/Fe]}$, $\mathrm{[Fe/H]}$ and age for the 351 \emph{TESS}-CVZ bright giants which have detailed abundances from APOGEE-DR16 and results from \texttt{PARAM} based on the BHM seismic parameters. The upper panel shows the $\mathrm{[Mg/Fe]}$-$\mathrm{[Fe/H]}$ plane coloured by stellar age. The dashed line indicates the visually defined separation which we make between high and low $\mathrm{[\alpha/Fe]}$ stars. The lower panels show each of these abundance ratios as a function of stellar age, and colour indicates there the mean orbital radius $R_{\mathrm{mean}}$ of each star. There is a clear dichotomy in age between the old, high $\mathrm{[Mg/Fe]}$ stars and the younger, low $\mathrm{[Mg/Fe]}$ stars. Similarly, the older populations have markedly lower mean radii and are on excursions from the inner regions of the disc. }
    \label{fig:mgfe_feh_age}
\end{figure*}

Further, \autoref{fig:mgfe_feh_age} demonstrates the relationship between age, element abundances and kinematics of the \emph{TESS}-SCVZ sample with APOGEE DR16 spectra (513 stars). The top panel shows the $\mathrm{[Fe/H]}$-$\mathrm{[Mg/Fe]}$ plane \citep[which has been studied extensively in a Galactic Archaeology context, e.g.][]{2013A&A...560A.109H,2014ApJ...796...38N,2015ApJ...808..132H,2020arXiv200712915Q} coloured by the stellar age (based on analysis using the more precise APOGEE DR16 stellar parameters). The high $\mathrm{[Mg/Fe]}$ stars which separate from the solar $\mathrm{[Mg/Fe]}$ population at $\mathrm{[Fe/H]} \lesssim 0.2$ are clearly separate in age, and are generally older than $\sim 10\ \mathrm{Gyr}$ (aside from some notable outliers, see below). The solar $\mathrm{[Mg/Fe]}$ population has a wider spread in ages below $\sim 10$ Gyr, and a relatively clear trend toward younger age with decreasing $\mathrm{[Mg/Fe]}$. These trends are also broadly seen in other studies of the $\mathrm{[Fe/H]}-\mathrm{[Mg/Fe]}$ plane with stellar ages \citep[or their proxies, e.g.][]{2013A&A...560A.109H,2016arXiv160407771A,2019ApJ...871..181H,2019MNRAS.489..176M}.

This trend is much clearer in the lower right panel, which shows $\mathrm{[Mg/Fe]}$ as a function of age, now coloured by the mean orbital radius for each star. As age increases from $\sim 0$ Gyr up to $\sim 10$ Gyr, the mean $\mathrm{[Mg/Fe]}$ increases by almost 0.1 dex. Stars older than $\sim 10$ Gyr have a larger spread in $\mathrm{[Mg/Fe]}$ and are those corresponding to the high $\mathrm{[Mg/Fe]}$ track in the upper panel. The $R_{\mathrm{mean}}$ of these stars indicates that they are likely to be members of populations formed in the inner disc of the Galaxy which have eccentric orbits and thus have travelled to the solar radius.

The mean orbital radii also indicate useful information in the lower left panel, showing $\mathrm{[Fe/H]}$ as a function of age, again coloured by $R_{\mathrm{mean}}$. At intermediate age ($\sim 3$ to $8$ Gyr) and, to a lesser extent outside of this range, there is a large spread in $\mathrm{[Fe/H]}$ at fixed age. Stars with high $R_{\mathrm{mean}}$ tend to be young and have lower $\mathrm{[Fe/H]}$. The most metal rich stars are old and have lower $R_{\mathrm{mean}}$. This gradient in $R_{\mathrm{mean}}$ across the range in $\mathrm{[Fe/H]}$ is to be expected given that the Milky Way disc has a shallow, yet significant, metallicity gradient \citep[e.g.][]{2012ApJ...746..149C,2014A&A...565A..89B,2016arXiv160804951A}. The lack of a clear relationship between age and $\mathrm{[Fe/H]}$ is likely a manifestation of radial migration in the disc, and has been discussed at length in the literature \citep{2002MNRAS.336..785S,2012MNRAS.422.1363S,2012A&A...548A.126M,2013A&A...558A...9M,2014ApJ...794..173V,2015MNRAS.447.4018G,2018ApJ...865...96F,2020arXiv200204622F}. 

\begin{figure}
    \centering
    \includegraphics[width=0.8\columnwidth]{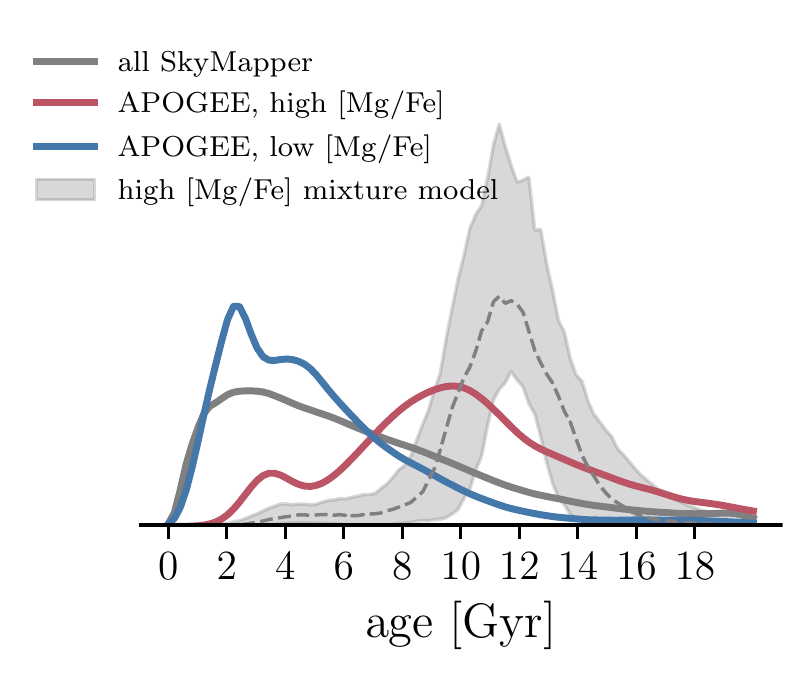}
    \caption{The posterior distributions of observed ages for high and low $\mathrm{[Mg/Fe]}$ stars in common with APOGEE (shown by the red and blue curves, respectively), and the full sample in common with \emph{SkyMapper} (in gray). We estimate the posterior using a Gaussian kernel density estimation applied to samples from the individual age posteriors from \texttt{PARAM}. The mixture model for the intrinsic age distribution of high $\mathrm{[Mg/Fe]}$ stars and its 1$\sigma$ uncertainty is shown by the gray dashed line and band. Despite the relatively large age uncertainties, the mixture model recovers the intrinsic age distribution for high $\mathrm{[Mg/Fe]}$ stars found by previous works.}
    \label{fig:agedist}
\end{figure}

Finally, we show the age distributions of our samples in \autoref{fig:agedist}. We divide the APOGEE based ages into high and low $\mathrm{[Mg/Fe]}$ groups, using the visually determined division shown in the top panel of \autoref{fig:mgfe_feh_age}. We estimate the posterior distribution of the observed ages using a Gaussian kernel density estimation applied to 1000 samples from the posterior on each star from \texttt{PARAM}. The \emph{SkyMapper} and low $\mathrm{[Mg/Fe]}$ APOGEE ages have qualitatively similar distributions, suggesting that the \emph{SkyMapper} sample is dominated by low $\mathrm{[Mg/Fe]}$ stars. The high $\mathrm{[Mg/Fe]}$ stars have a bimodal age distribution, with a strong peak at $\sim 10$ Gyr. This peak is consistent with the old age for the high $\mathrm{[\alpha/Fe]}$ disc as shown by a number of studies \citep[e.g.][]{2011MNRAS.414.2893F,2013A&A...560A.109H,2018MNRAS.475.5487S,2020arXiv200414806M}. The young peak is likely to be a result of `over-massive', apparently `young' $\alpha$-rich stars, contaminating the sample \citep{2015MNRAS.451.2230M,2015A&A...576L..12C,2016A&A...595A..60J,2019MNRAS.487.4343H}. These stars are outliers to the true age distibution of the high $\mathrm{[\alpha/Fe]}$ disc. Evidence collected thus far strongly suggests that these are stars which have undergone some mass transfer with a (now unseen) companion earlier in their lives and thus appear young to asteroseismic methods \citep[see discussion in, e.g.][]{2020arXiv200414806M}. Each sample has a long tail extending to high ages driven by the uncertainties on individual ages, which are Gaussian in $\log_{10}(\mathrm{age})$, due to the prior adopted in \texttt{PARAM}. 

We fit a mixture model for the intrinsic age distribution of the high $\mathrm{[Mg/Fe]}$ stars, following the methodology of \citet{2020arXiv200414806M}. This model accounts for the contamination by `over-massive' stars, and recovers the intrinsic spread in age of the population (i.e. other than that caused by the age uncertainties). We find results that are consistent with those of \citet{2020arXiv200414806M}: the mean age of the population is $\mu = 11.3\pm 0.8\ \mathrm{Gyr}$, with an intrinsic spread (defined as the standard deviation in linear age) of $\delta_{\mathrm{age}} = 1.3^{+0.6}_{-0.8}\ \mathrm{Gyr}$. The best fit model and the 1$\sigma$ uncertainty are shown by the dashed gray line and band in \autoref{fig:agedist}. It is clear that the age uncertainties inflate the distribution considerably, but careful modelling accounting for these uncertainties can recover the underlying age spread. This demonstrates that it will still be possible to make important inferences of, for example, the Galactic star formation history using \emph{TESS} ages, despite the relatively lower asteroseismic precision (relative to, e.g. \emph{Kepler}) afforded by the data. 

Finally, we note that our mixture model finds a contamination of the high $\mathrm{[Mg/Fe]}$ stars by `over-massive' stars at a level of $15^{+11}_{-7} \%$. This contamination is a little higher but not inconsistent above the $1\sigma$ level, than the results of \citet{2020arXiv200414806M}. It is likely that the differences in selection between these two works likely brings about some population bias which may affect these numbers. Future studies which aim to determine the origin and nature of these `over-massive' stars will need to account for these biases.

\section{Summary \& Conclusions} \label{sec:conclusions}

We have demonstrated the strong potential for Galactic archaeology and stellar astrophysics and asteroseismic quality of year-long photometry from the \emph{TESS}-CVZ. Studying a sample of $15,405$ giants in- and outside the CVZ, we demonstrate that currently achieved seismic yields are comparable or better to those estimated using a model of detection probability in \emph{TESS} and produce precise stellar ages, masses and radii when combined with spectroscopic constraints. Power spectra derived from the time series data provide clear detections of oscillation modes in stars in a large range of evolutionary states, providing great potential for new constraints on stellar evolution models. To summarise our findings:
\begin{itemize}
    \item \textbf{Systematics:} A number of systematics are evident in the light curves that we generate based on the \emph{TESS} FFI, which appear worsened in the 1 year data. More detailed correction of these will be required to ascertain the best possible seismic data. It is likely that many such issues can be resolved with some extra treatment of our light curves.
    \item \textbf{Seismic detection yields:} We find that the realised detection yields from the \emph{TESS}-CVZ compare well with a simple model for the detection probability, but only after accounting for dilution effects. After confirming seismic detections using luminosities derived with 2MASS and \emph{Gaia} for 8,249 stars, we find an average detection yield of $\sim 36\%$ (2,890 stars), which increases to $\sim50\%$ when stars with a year of data are considered. We presently suffer from issues in procuring seismic parameters for brighter stars with $\nu_{\mathrm{max}} < 10\ \mathrm{\mu Hz}$, which likely will require specifically tuned analysis methods.
    \item \textbf{Constraints on stellar mass and age: } we produce mass and age constraints for 1,749 stars, based on \emph{SkyMapper}/\emph{TESS}-HERMES stellar parameters and BHM seismic parameters using the \texttt{PARAM} Bayesian parameter estimation code. By comparing the uncertainties on mass and age when \emph{Gaia} luminosities are used as constraints or not, we show that without refining the current seismic parameter measurmements, an external luminosity constraint is required to achieve adequate precision (i.e. less than 10\% in mass and 30\% in age) on these parameters.
    \item \textbf{Seismology across the Hertzsprung-Russell diagram: } The \emph{TESS}-CVZ data provides good power spectral resolution even for relatively low $\nu_{\mathrm{max}}$ stars. We show that stars on the AGB bump are clearly detected in our sample, and the detection of mixed modes and, tentatively, rotational splitting is possible in RC stars with 12 sectors of data. Importantly, target selection effects are more simply accounted for in the \emph{TESS}-FFI samples than in, for example, \emph{Kepler}. We demonstrate that our data compare well with \emph{Kepler} data analysed in similar ways by showing  
    \item \textbf{Galactic archaeology in the Milky Way disc: } The \emph{TESS}-SCVZ sample provides a good sampling of velocity space in the solar vicinity. Trends between stellar mass (and therefore age) and the mean radii and maximum vertical excursions of stellar orbits are clearly visible. By integrating the orbits of stars in our sample, we demonstrate that the data from the CVZ, although limited in spatial extent, `dynamically' samples a large section of the Milky Way disc with $5 \lesssim R \lesssim 10$ kpc. The age precision afforded by \emph{TESS}, in combination with high resolution spectroscopic data from APOGEE, is enough to distinguish clear differences in age between the high and low $\mathrm{[Mg/Fe]}$ disc stars and will be of great utility in properly reconstructing the star formation history of the disc. 
\end{itemize}
In particular, this paper has demonstrated the great importance of spectroscopic follow up for giants in the \emph{TESS}-CVZs. We predict that seismic samples of the order of a few $10^{4}$ giants could be attainable in the CVZs, but age and mass estimates will only be readily available -- and sufficiently precise -- for stars which also have spectroscopic information. Such spectroscopic data will likely become available from future survey data (e.g. GALAH: \citeauthor{2016arXiv160902822M} \citeyear{2016arXiv160902822M}  SDSS-V: \citeauthor{2019BAAS...51g.274K} \citeyear{2019BAAS...51g.274K}), but high resolution, high SNR samples may be of great utility to the community as calibration data.

Careful modelling of large samples of asteroseismic data will afford detailed insights into the formation and evolution in the disc. Furthermore, it is likely possible to further pin down the precision on these stellar ages, using novel analysis techniques based on machine learning \citep{2020arXiv200509682B,2018MNRAS.476.3233H,2018ApJ...866...15N} or more detailed modelling (and fitting) of the power spectra \citep[e.g.][]{2019MNRAS.484..771R,2020arXiv200601783M}. In addition to this, it is foreseeable that samples from the CVZs will be useful as training data for future data-driven analyses, providing time-series with consistent noise statistics but longer baselines than the whole-sky \emph{TESS} data. 

\section*{Acknowledgements}

JTM and AM acknowledge support from the ERC Consolidator Grant funding scheme (project ASTEROCHRONOMETRY, G.A. n. 772293). JTM acknowledges support from the Banting Postdoctoral Fellowship programme administered by the Government of Canada, and a CITA/Dunlap Institute fellowship. The Dunlap Institute is funded through an endowment established by the David Dunlap family and the University of Toronto. S.M.\ acknowledges support from the Spanish Ministry with the Ramon y Cajal fellowship number RYC-2015-17697. RAG acknowledges the support from the PLATO CNES grant. LC acknowledges support from the Australian Research Council grants FT160100402. TC acknowledges support from the European Union's Horizon 2020 research and innovation programme under the Marie Sk\l{}odowska-Curie grant agreement No.~792848 (PULSATION). A.S. is partially supported by grants ESP2017-82674-R (Spanish Government) and 2017-SGR-1131 (Generalitat de Catalunya). MHP and MV acknowledge support from NASA grant 80NSSC18K1582.

This paper includes data collected by the \emph{TESS} mission. Funding for the \emph{TESS} mission is provided by the NASA Explorer Program.

This work has made use of data from the European Space Agency (ESA) mission {\it Gaia} (\url{https://www.cosmos.esa.int/gaia}), processed by the {\it Gaia} Data Processing and Analysis Consortium (DPAC, \url{https://www.cosmos.esa.int/web/gaia/dpac/consortium}). Funding for the DPAC has been provided by national institutions, in particular the institutions participating in the {\it Gaia} Multilateral Agreement.  

Funding for the Sloan Digital Sky Survey IV has been provided by the Alfred P. Sloan Foundation, the U.S. Department of Energy Office of Science, and the Participating Institutions. SDSS-IV acknowledges support and resources from the Center for High-Performance Computing at the University of Utah. The SDSS web site is www.sdss.org. SDSS-IV is managed by the Astrophysical Research Consortium for the Participating Institutions of the SDSS Collaboration including the Brazilian Participation Group, the Carnegie Institution for Science, Carnegie Mellon University, the Chilean Participation Group, the French Participation Group, Harvard-Smithsonian Center for Astrophysics, Instituto de Astrof\'isica de Canarias, The Johns Hopkins University, Kavli Institute for the Physics and Mathematics of the Universe (IPMU) / University of Tokyo, Lawrence Berkeley National Laboratory, Leibniz Institut f\"ur Astrophysik Potsdam (AIP), Max-Planck-Institut f\"ur Astronomie (MPIA Heidelberg), Max-Planck-Institut f\"ur Astrophysik (MPA Garching), Max-Planck-Institut f\"ur Extraterrestrische Physik (MPE), National Astronomical Observatories of China, New Mexico State University, New York University, University of Notre Dame, Observat\'ario Nacional / MCTI, The Ohio State University, Pennsylvania State University, Shanghai Astronomical Observatory, United Kingdom Participation Group, Universidad Nacional Aut\'onoma de M\'exico, University of Arizona, University of Colorado Boulder, University of Oxford, University of Portsmouth, University of Utah, University of Virginia, University of Washington, University of Wisconsin, Vanderbilt University, and Yale University.  

This research made use of matplotlib, a Python library for publication quality graphics \citep{Hunter:2007}, NumPy \citep{van2011numpy}, SciPy \citep{Virtanen_2020}, the IPython package \citep{PER-GRA:2007}, Scikit-learn \citep{scikit-learn}, Astropy, a community-developed core Python package for Astronomy \citep{2018AJ....156..123A, 2013A&A...558A..33A}, Lightkurve, a Python package for Kepler and \emph{TESS} data analysis \citep{2018ascl.soft12013L} and Astroquery \citep{2019AJ....157...98G}.

\section*{Data Availability}

The processed data underlying this article are available in the article and in its online supplementary material. The raw data (e.g. unprocessed light curves, original data tables) can be made available on reasonable request to the corresponding author.




\bibliographystyle{mnras}
\bibliography{bib} 




\appendix

\appendix

\section{Surface gravity constraints for stars with missing data}
\label{app:A}

For the best possible constraints on stellar mass and age using a method such as \texttt{PARAM}, we require as many constraints on other fundamental stellar parameters as possible. At present, the TESS-SCVZ data are lacking in publicly available spectroscopic follow up, and so these stellar parameters must be gathered by alternative means. As described above, we use a combination of data from \emph{SkyMapper}, \emph{TESS}-HERMES, and APOGEE. There are a number of stars for which an estimate of $\log(g)$ is not available from \emph{TESS}-HERMES. Here we exploit the fact that there is a correlation between $T_{\mathrm{eff}}$ and luminosity $L$, which in turn correlates with $R$ and thus the surface gravity $\log(g)$ with some dependence on $\mathrm{[Fe/H]}$, to constrain a prior on $\log(g)$ for stars which have $T_{\mathrm{eff}}$ and $\mathrm{[Fe/H]}$. 

\begin{figure}
    \centering
    \includegraphics[width=0.8\columnwidth]{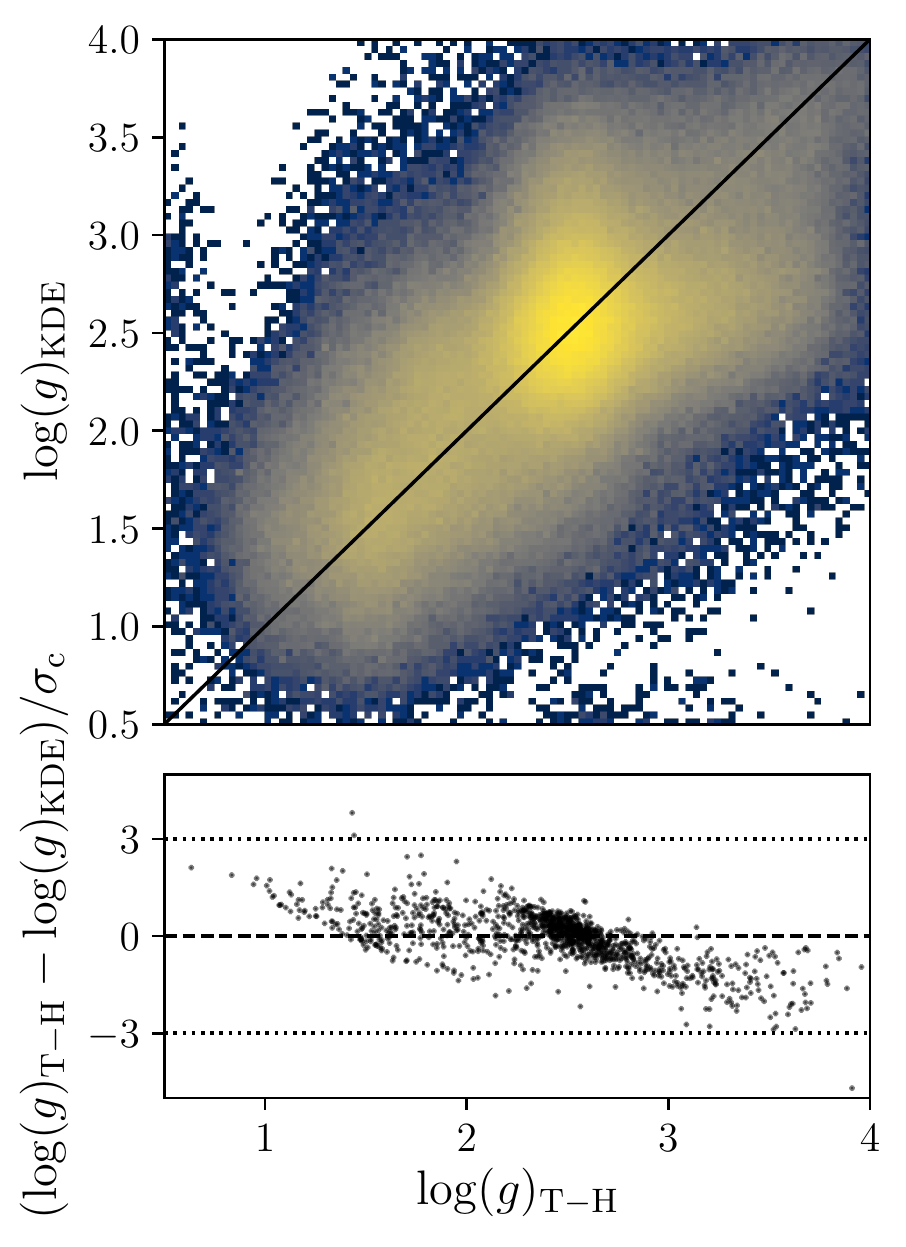}
    \caption{Comparison of $\log(g)$ from \emph{TESS}-HERMES with that inferred using the \emph{TESS}-HERMES and \emph{SkyMapper} data as prior information. The top panel shows the density distribution of samples from the posterior. when the \emph{TESS}-HERMES $log{g}$ is used, and when the $\log(g)$ is inferred from the prior. The bottom panel compares the difference between the inferred and measurred $\log(g)$, normalised by the combined uncertainty $\sigma_c$. $\sim93\%$ of the inferences of $\log(g)$ fall within $1\sigma$ of the measured values, although the uncertainty on the inferred $\log(g)$ is a factor of 3 greater than those measured by \emph{TESS}-HERMES.}
    \label{fig:loggs}
\end{figure}

We use a simple method to infer $\log(g)$ for those stars which only had photometric $T_{\mathrm{eff}}$ and $\mathrm{[Fe/H]}$ from \emph{SkyMapper}. Using the 1186 stars in our sample for which spectroscopic $\log(g)$ \emph{was} available from \emph{TESS}-HERMES, we construct a Gaussian KDE of $\log(g)$, $T_{\mathrm{eff}}$ and $\mathrm{[Fe/H]}$ space, adopting a kernel with a bandwidth determined via Scott's rule, as implemented in \texttt{scipy}. The resulting KDE is smooth in all three parameters when marginalising over the range of the other parameters, suggesting it provides a fair representation of the prior information on the $\log(g)$ distribution. By combining the prior information on $T_{\mathrm{eff}}$ and $\mathrm{[Fe/H]}$ for each star (assuming Gaussian uncertainties) with this KDE, we construct the joint posterior probability for all three parameters:
\begin{multline}
    \ln p(\log(g),T_{\mathrm{eff}},\mathrm{[Fe/H]}) = \ln p(\mathrm{KDE}(\log(g),T_{\mathrm{eff}},\mathrm{[Fe/H]})) \\ + \ln p(T_{\mathrm{eff}}) + \ln p(\mathrm{[Fe/H]}),
\end{multline}
where $\mathrm{KDE}(\log(g),T_{\mathrm{eff}},\mathrm{[Fe/H]})$ is the density of the KDE at a given set of parameters, and $p(T_{\mathrm{eff}})$,  $p(\mathrm{[Fe/H]})$ are the prior information on $T_{\mathrm{eff}}$ and $\mathrm{[Fe/H]}$ from \emph{SkyMapper}. For each star with missing $\log(g)$, we use rejection sampling to take samples of this posterior within the range of parameters represented in the full data set. We verify that this procedure gives a reliable constraint on $\log(g)$ by making an inference of the parameter for the set of stars with \emph{TESS}-HERMES $\log(g)$ constraints, and find that the \emph{SkyMapper} prior-based $\log(g)$ constraint is within $3\sigma$ of the \emph{TESS}-HERMES measurement for all 1186 stars, and within $1\sigma$ for $\sim 93\%$ of the sample. This suggests that this inference is reliable for the rest of the dataset, and the resulting uncertainty on the $\log(g)$ constraint is realistic and will hence be propagated into our inference of luminosity, mass, age and radii consistently. \autoref{fig:loggs} shows the comparison between our inferred $\log(g)$ and that from \emph{TESS}-HERMES. The median uncertainty on the inferred $\log(g)$ is 0.3, compared to 0.1 for the \emph{TESS}-HERMES measurements.

\section{\emph{Gaia} parallax zero-point offset}
\label{app:C}

Since this sample represents an extension to brighter magnitudes of the set of stars with asteroseismic constraints, in a different region of the sky to previously available data (e.g. \emph{Kepler}), it is appropriate to check if and how the derived parallax zero-point offset may change. While this is not the focus of this work and will likely form the basis of future, more detailed, studies based on \emph{TESS} data, we make a simplified assessment of this offset here. Current assessments of the parallax zero-point offset in \emph{Gaia} DR2 yield values of $30 \lesssim \Delta\varpi \lesssim 60\ \mathrm{\mu as}$  \citep[e.g.][]{2018A&A...616A...2L,2019MNRAS.487.3568S,2019A&A...628A..35K,2019MNRAS.486.3569H,2019arXiv190208634L,2019ApJ...878..136Z,2020MNRAS.493.4367C}. Furthermore, it has been widely shown that the zero-point is multivariate as a function of position on the sky \citep[e.g.][]{2018A&A...616A...2L,2020MNRAS.493.4367C}, and stellar parameters \citep[e.g.][]{2019ApJ...878..136Z,2019A&A...628A..35K}. 

Here, we make a simple assessment of the zero-point offset for the \emph{TESS}-SCVZ sample only as a function of the apparent $K_S$ band magnitude. We take the raw \emph{Gaia} DR2 parallax and its associated uncertainty, and derive a `seismic' parallax by transforming the \texttt{PARAM} $M_{bol}$ values, estimated using the BHM seismic parameters and the \emph{SkyMapper} stellar parameters, using the bolometric corrections derived for the sample based also on these parameters. We use a separate run of \texttt{PARAM} which did not use the \emph{Gaia} derived luminosity as a constraint such that the luminosity is directly predicted by the seismic and stellar parameters. We then sample the resultant posterior on $\Delta\varpi = \varpi_{\mathrm{PARAM}} - \varpi_{Gaia}$ and $K_S$ for each star and use the $0.16,0.5$ and $0.84$ quartiles as input to our fitting procedure. We use the likelihood function for a linear fit with two dimensional uncertainties (assuming that the uncertainty on $\Delta\varpi$ and $K_S$ are uncorrelated) given in equation (32) of \citet{2010arXiv1008.4686H}, and sample the posterior distribution of the parameters describing a linear relationship between $\Delta\varpi$ and $K_S$ using \texttt{emcee} \citep{2013PASP..125..306F}. 

\begin{figure}
    \centering
    \includegraphics[width=\columnwidth]{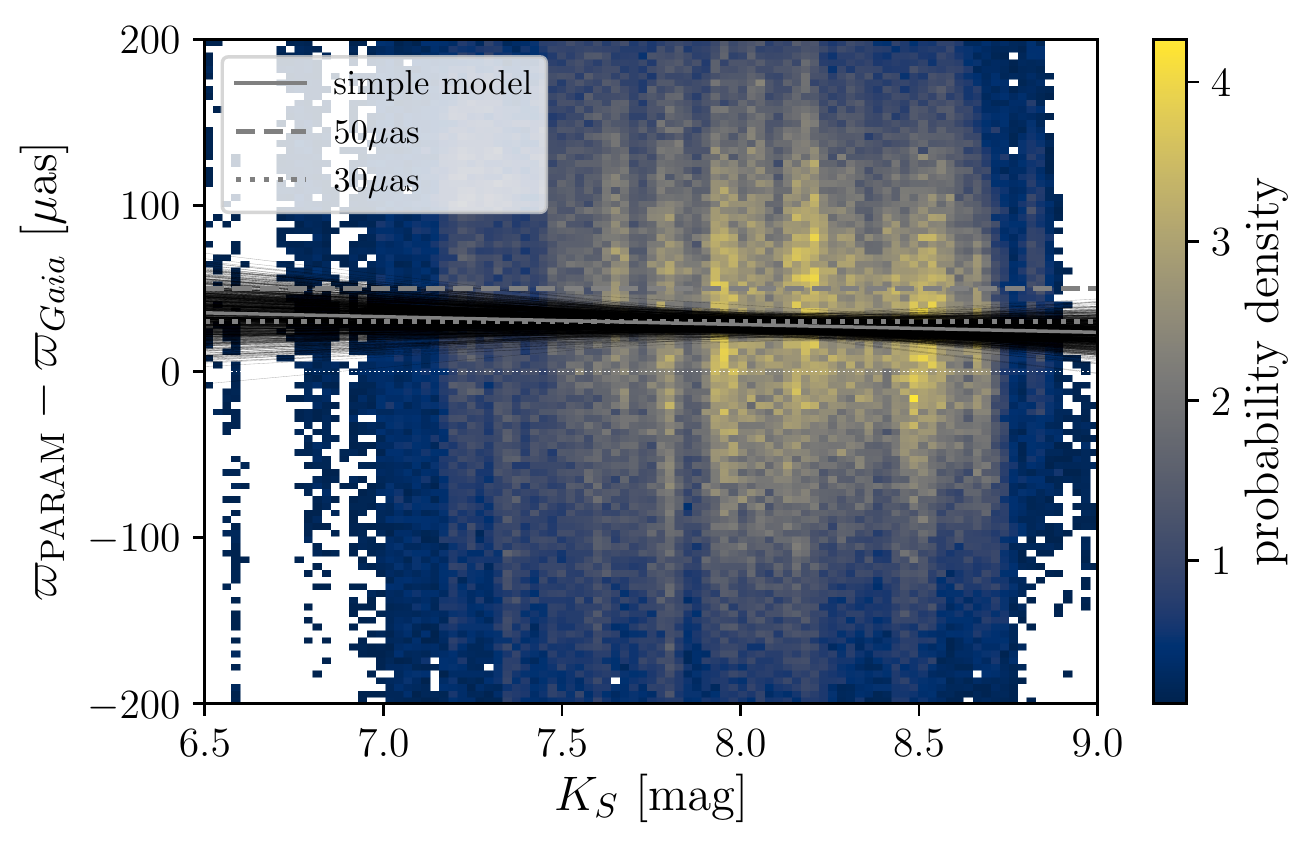}
    \caption{The \emph{Gaia} DR2 parallax zero-point offset as a function of $K_S$ magnitude, as measured by comparing the observed \emph{Gaia} parallax values with those derived by measuring stellar luminosities using the BHM seismic results and PARAM (removing the constraint based on the \emph{Gaia} luminosity).  The 2D histogram shows the density of samples from the posteriors on magnitude and parallaxes, assuming Gaussian uncertainties. The dotted and dashed lines demonstrate the levels of common values for the zero-point offset measured using various techniques. Our best fit model is shown by the solid gray line, and MCMC samples of the posterior likelihood function of the parameters of the fit given the data is shown by the individual black lines. The best fit model gives a dependency of the offset on $K_S$ of $\Delta\varpi = (-5\pm6)(K_S-8.) + (30\pm3)\ \mathrm{\mu as}$.} 
    \label{fig:parallax}
\end{figure}

The resulting fit and posterior samples compared to the data are shown in \autoref{fig:parallax}. The data are shown as a two-dimensional histogram of the samples from the posteriors on $\Delta\varpi$ and $K_S$, to give a sense of the large uncertainties on $\Delta\varpi$. Despite these large uncertainties, an offset of the mean $\Delta\varpi$ from 0 is clear. We show offsets of $30$ and $50\ \mathrm{\mu as}$ as dotted and dashed lines respectively, as a guide to how our derived offset compares with those from the literature. Our best fit model is consistent with being flat as a function of $K_S$, with a zero-point offset of $\sim 30\ \mathrm{\mu as}$, such that $\Delta\varpi = (-5\pm6)(K_S-8.) + (30\pm3)\ \mathrm{\mu as}$. This offset is somewhat smaller than many derived using currently available asteroseismic samples \citep[e.g.][]{2019MNRAS.486.3569H,2019A&A...628A..35K,2019ApJ...878..136Z}. This may be a genuine population effect, due to the fact that this sample is somewhat closer and thus brighter than the \emph{Kepler} sample. This may also reflect noted variations as a function of position on the sky. Future studies of the \emph{TESS} data will likely narrow down this estimation and trends with these parameters.

\section{Comparing photometric and asteroseismic luminosities}
\label{app:B}

\begin{figure*}
    \centering
    \includegraphics[width=\textwidth]{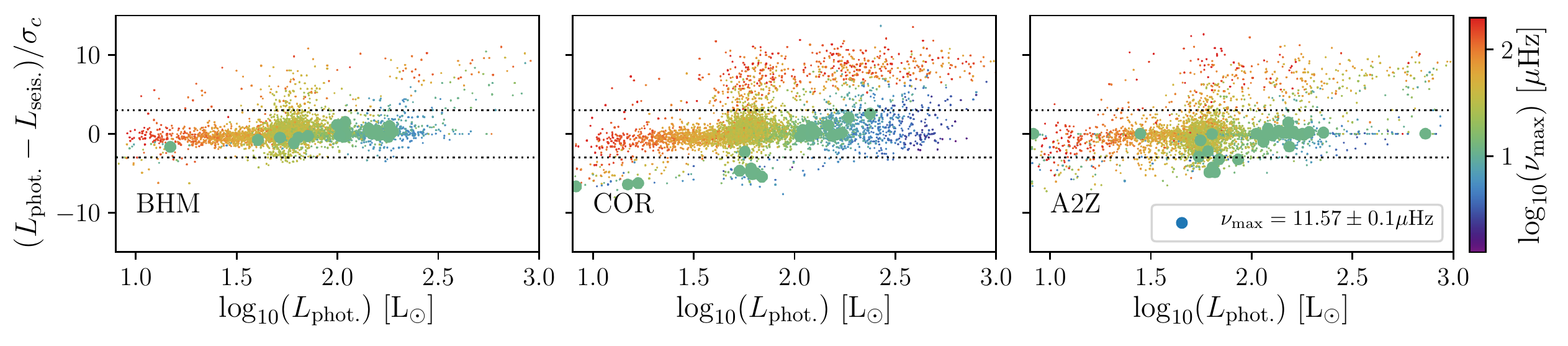}
    \caption{Comparison between photometrically and seismically derived luminosities. The seismic luminosities $L_{\mathrm{seis.}}$ are estimated using the scaling relations applied to the seismic parameters determined by the BHM, COR and A2Z pipelines, from left to right. The vertical axes give the difference between photometric and seismic luminosity, normalised by the combined uncertainty $\sigma_c$. The dashed horizontal lines demonstrate the range at which results are consistent within $3\sigma_{c}$. The majority of stars have consistent luminosities in all three pipelines, but each have clear cases of false positive detections. We highlight stars with $\nu_{\mathrm{max}}$ near to the diurnal frequency ($11.57 \mathrm{\mu Hz}$ as larger points, demonstrating that most detections at this frequency appear to be true positives, and not due to systematic effects in our light curves.} 
    \label{fig:photcomp}
\end{figure*}


\autoref{fig:photcomp} shows the full comparison between $L_{\mathrm{phot.}}$ and $L_{\mathrm{seis.}}$ for each of the pipelines considered. We normalise the difference between $L_{\mathrm{phot.}}$ and $L_{\mathrm{seis.}}$ by the combined uncertainty on photometric and seismic luminosities $\sigma_{c}$. In this way, a difference of e.g. $3\sigma_c$ indicates the two measurements are consistent to $3\sigma$. The uncertainty on $L_{\mathrm{seis.}}$ is considerably higher than that on $L_{\mathrm{phot.}}$ in many cases, which can lead to small values of this difference. However, since this indicates the uncertainties are well estimated, we allow these into our sample. The horizontal lines in each figure show the $\pm 3\sigma_{c}$ limit. For each pipeline, we colour the points by $\nu_\mathrm{max}$ and show (as larger points) the stars for which each pipeline returned a $\nu_{\mathrm{max}}$ close to the diurnal frequency ($\nu_\mathrm{max} = 11.57\pm 0.1$). This will allow us to assess whether the appearance of these detections is real or due to some systematic in the \emph{TESS} photometry. 

Inspection of \autoref{fig:photcomp} reveals that there are a considerable number of targets across all three pipelines for which $L_{\mathrm{seis.}}$ is not consistent with $L_{\mathrm{phot.}}$ (within $3\sigma_{c}$). In each pipeline, a cloud of `high' $\nu_\mathrm{max} \gtrsim 10^2$ detections is visible such that $L_\mathrm{seis.}$ underestimated by $\simeq 10\sigma_{c}$. This means that many intrinsically bright targets have been misidentified by the seismic analysis as high $\nu_{\mathrm{max}}$ and therefore low luminosity stars. Visual inspection of the power spectra of a few of these targets shows that it appears that there is genuine power excess at low $\nu$ as expected from their luminosity. However, a number of features are apparent in the noisier, high $\nu$ part of the power spectrum. This likely indicates some fine-tuning of analysis methods may be necessary for \emph{TESS} power spectra.

Furthermore, it appears that there are a number of genuine and photometrically consistent detections with $\nu_\mathrm{max}\simeq11.57\ \mathrm{\mu\ Hz}$, suggesting that systematics at this freqeuncy are not as problematic as may be expected, given the scattered light present in many of the \emph{TESS} observations. This also suggests that an apparent `bump' in the $\nu_\mathrm{max}$ distribution at roughly this frequency may be a real feature.


\section{Catalogue of seismic, photometric, spectroscopic and kinematic properties for \emph{TESS}-SCVZ giants}

\label{app:D}

We provide all data necessary for the reproduction of the analyses presented here in a catalogue of the compiled seismic, photometric, spectroscopic and kinematic properties for the 15405 stars in the \emph{TESS}-SCVZ bright red giant sample. The columns included in the data table are described in \autoref{tab:datamodel}. All relevant columns described in \autoref{tab:datamodel} have associated uncertainties, which are adjacent to those columns in the table. The catalogue reproduces data produced by the \emph{Gaia} \citep{2018arXiv180409365G}, 2MASS \citep{2006AJ....131.1163S}, \emph{TESS}-HERMES \citep{2018MNRAS.473.2004S}, \emph{SkyMapper} \citep{2019PASA...36...33O} and APOGEE \citep{2015arXiv150905420M} surveys, and includes orbital parameters estimated as described in \autoref{sec:data}.

Since we include all the possible data in the catalogue (i.e. we do not remove stars which we flagged as bad in any way), we recommend that users take care to apply the appropriate selection to any data that is used. 
For example, we recommend that when using PARAM results, the sample is limited to only those stars with stellar radii measured to be less than $12 R_{\odot}$ \citep[as recommended by][]{2020arXiv200414806M} and those with the appropriate luminosity flag indicating that the seismic parameters are robust relative to the \emph{Gaia} luminosity. 
Furthermore, since the internal consistency is not optimal in this first set of data, we recommend that the mean $\nu_{\mathrm{max}}$ and $\Delta\nu$ values are avoided in most cases, as these can represent large deviations from the true values in some cases. 
Best practice might be to use a set of results from a single pipeline, restricting the sample to only those stars where consistent detections were made across all three pipelines (i.e. \texttt{numax\_dnu\_consistent} $= 1$), and where the relevant luminosity consistency flag is on (i.e. \texttt{lum\_flag\_XXX} $= 1$, where \texttt{XXX} is the relevant pipeline identifier). In the case of the BHM results, this provides a sample of 1,749 stars which likely have the most robust data provided by that pipeline.

{\red As an additional way to assess the data quality for individual targets, we include an estimate of the goodness of fit in \texttt{PARAM} through the $\chi^{2}$ of the fit for those stars which returned parameter values. This value is computed by comparing the input constraints with those predicted by the best fit model. In this way, a star which is not well represented by the grid of stellar models employed by \texttt{PARAM} will have a high $\chi^{2}$. We recommend that users check samples for high values of this parameter, especially when using small numbers of stars. Any $\chi^{2}$ well above $\sim 4$ (the number of degrees of freedom) can be considered to be a poor fit, and likely represent a star whose parameters are not reproduced by the models. Approximately 90\% of the targets with $\chi^{2} > 4$ are removed by the flags outlined above. In the catalogue, these numbers are found under \texttt{CHI2\_gof\_PARAM\_BHM} and \texttt{CHI2\_gof\_APO\_PARAM\_BHM} for the results using \emph{SkyMapper} and APOGEE, respectively.}

\begin{table*}
\caption{The data model of the catalogue of \emph{TESS}-SCVZ targets which we release in this work. We compile seismic, photometric, spectroscopic and kinematic properties from \emph{Gaia} \citep{2018arXiv180409365G}, 2MASS \citep{2006AJ....131.1163S}, \emph{TESS}-HERMES \citep{2018MNRAS.473.2004S}, \emph{SkyMapper} \citep{2019PASA...36...33O} and APOGEE \citep{2015arXiv150905420M}. Orbital parameters are computed using the method presented in \citet{2018arXiv180202592M}, implemented in \texttt{galpy} \citep{2015ApJS..216...29B}. All relevant columns are accompanied by an associated uncertainty, defined either as the standard deviation or the 0.16 and 0.84 quantiles. Uncertainties are found in accompanying columns labeled with the suffix `\texttt{err}'. \label{tab:datamodel}}
\begin{tabular}{llc}
\hline
Column Identifier & Description & Units \\
\hline
\texttt{source\_id} & Gaia DR2 Source ID & $\mathrm{None}$ \\
\texttt{N\_sectors} & Number of sectors star was observed in & $\mathrm{27\ days}$ \\
\texttt{ra} & Right Ascension & $\mathrm{deg}$ \\
\texttt{dec} & Declination & $\mathrm{deg}$ \\
\texttt{l} & Galactic Longitude & $\mathrm{deg}$ \\
\texttt{b} & Galactic Latitude & $\mathrm{deg}$ \\
\texttt{ecl\_lon} & Ecliptic Longitude & $\mathrm{deg}$ \\
\texttt{ecl\_lat} & Ecliptic Latitude & $\mathrm{deg}$ \\
\texttt{parallax} & Gaia DR2 Parallax & $\mathrm{mas}$ \\
\texttt{pmra} & Gaia DR2 proper motion in RA & $\mathrm{mas\ yr^{-1}}$ \\
\texttt{pmdec} & Gaia DR2 proper motion in Dec & $\mathrm{mas\ yr^{-1}}$ \\
\texttt{radial\_velocity} & Gaia DR2 heliocentric radial velocity & $\mathrm{km\ s^{-1}}$ \\
\texttt{hmag} & 2MASS $H$-band magnitude & $\mathrm{mag}$ \\
\texttt{jmag} & 2MASS $J$-band magnitude & $\mathrm{mag}$ \\
\texttt{kmag} & 2MASS $K_{S}$-band magnitude & $\mathrm{mag}$ \\
\texttt{phot\_g\_mean\_mag} & Gaia DR2 $G$-band magnitude & $\mathrm{mag}$ \\
\texttt{phot\_bp\_mean\_mag} & Gaia DR2 $G_{BP}$-band magnitude & $\mathrm{mag}$ \\
\texttt{phot\_rp\_mean\_mag} & Gaia DR2 $G_{RP}$-band magnitude & $\mathrm{mag}$ \\
\texttt{numax\_COR} & $\nu_{\mathrm{max}}$ from COR pipeline & $\mathrm{\mu Hz}$ \\
\texttt{dnu\_COR} & $\Delta\nu$ from COR pipeline & $\mathrm{\mu Hz}$ \\
\texttt{numax\_BHM} & $\nu_{\mathrm{max}}$ from BHM pipeline & $\mathrm{\mu Hz}$ \\
\texttt{dnu\_BHM} & $\Delta\nu$ from BHM pipeline & $\mathrm{\mu Hz}$ \\
\texttt{numax\_A2Z} & $\nu_{\mathrm{max}}$ from A2Z pipeline & $\mathrm{\mu Hz}$ \\
\texttt{dnu\_A2Z} & $\Delta\nu$ from A2Z pipeline & $\mathrm{\mu Hz}$ \\
\texttt{mean\_numax} & Mean $\nu_{\mathrm{max}}$ between all pipelines returning results & $\mathrm{\mu Hz}$ \\
\texttt{mean\_dnu} & Mean $\Delta\nu$ between all pipelines returning results & $\mathrm{\mu Hz}$ \\
\texttt{N\_pipelines\_mean} & Number of pipelines included in mean & $\mathrm{None}$ \\
\texttt{seismic\_param\_gold} & Flag indicating all pipeline results for $\nu_{\mathrm{max}}$ and $\Delta\nu$ within $1\sigma$ of global mean & $\mathrm{boolean}$ \\
\texttt{logg\_HERMES} & Surface gravity from \emph{TESS}-HERMES & $\mathrm{None}$ \\
\texttt{Teff\_SKYMAPPER} & $T_{\mathrm{eff}}$ from \emph{SkyMapper} & $\mathrm{K}$ \\
\texttt{feh\_SKYMAPPER} & $\mathrm{[Fe/H]}$ from \emph{SkyMapper} & $\mathrm{None}$ \\
\texttt{ecc\_MB2018} & Orbit eccentricity in \texttt{MWPotential2014} & $\mathrm{None}$ \\
\texttt{rperi\_MB2018} & Pericenter radius in \texttt{MWPotential2014} & $\mathrm{kpc}$ \\
\texttt{rap\_MB2018} & Apocenter radius in \texttt{MWPotential2014} & $\mathrm{kpc}$ \\
\texttt{zmax\_MB2018} & Maximum vertical excursion in \texttt{MWPotential2014} & $\mathrm{kpc}$ \\
\texttt{APOGEE\_ID\_APOGEE} & APOGEE ID in DR16 & $\mathrm{None}$ \\
\texttt{FE\_H\_APOGEE} & APOGEE $\mathrm{[Fe/H]}$ (DR16) & $\mathrm{None}$ \\
\texttt{MG\_FE\_APOGEE} & APOGEE $\mathrm{[Mg/Fe]}$ (DR16) & $\mathrm{None}$ \\
\texttt{LOGG\_APOGEE} & APOGEE $\log(g)$ (DR16) & $\mathrm{None}$ \\
\texttt{TEFF\_APOGEE} & APOGEE $T_{\mathrm{eff}}$ (DR16) & $\mathrm{K}$ \\
\texttt{age\_PARAM\_BHM} & Age from PARAM, based on BHM and \emph{SkyMapper} & $\mathrm{Gyr}$ \\
\texttt{mass\_PARAM\_BHM} & Mass from PARAM, based on BHM and \emph{SkyMapper} & $\mathrm{M_{\odot}}$ \\
\texttt{rad\_PARAM\_BHM} & Radius from PARAM, based on BHM and \emph{SkyMapper} & $\mathrm{R_{\odot}}$ \\
\texttt{CHI2\_gof\_PARAM\_BHM} & $\chi^{2}$ value from \texttt{PARAM} computed based on distance between input constraints and best fit model parameters \\
\texttt{mbol\_PARAM\_BHM\_NO\_L} & Bolometric magnitude from PARAM, based on BHM and \emph{SkyMapper} without using Luminosity as a constraint & $\mathrm{mag}$ \\
\texttt{age\_APO\_PARAM\_BHM} & Age from PARAM, based on BHM and APOGEE & $\mathrm{Gyr}$ \\
\texttt{mass\_APO\_PARAM\_BHM} & Mass from PARAM, based on BHM and APOGEE & $\mathrm{M_{\odot}}$ \\
\texttt{rad\_APO\_PARAM\_BHM} & Radius from PARAM, based on BHM and APOGEE & $\mathrm{R_{\odot}}$ \\
\texttt{CHI2\_gof\_APO\_PARAM\_BHM} & $\chi^{2}$ value from \texttt{PARAM} computed based on distance between input constraints and best fit model parameters \\
\texttt{luminosity\_BHM} & Luminosity from seismic scaling relations applied to BHM & $\mathrm{L_{\odot}}$ \\
\texttt{luminosity\_COR} & Luminosity from seismic scaling relations applied to COR & $\mathrm{L_{\odot}}$ \\
\texttt{luminosity\_A2Z} & Luminosity from seismic scaling relations applied to A2Z & $\mathrm{L_{\odot}}$ \\
\texttt{luminosity\_HERMES} & Luminosity based on 2MASS photometry and bolometric correction based on \emph{TESS}-HERMES & $\mathrm{L_{\odot}}$ \\
\texttt{luminosity\_GAIA} & Luminosity based on 2MASS photometry and bolometric correction based on \emph{SkyMapper} and \emph{Gaia} DR2 parallax & $\mathrm{L_{\odot}}$ \\
\texttt{luminosity\_APO\_GAIA} & Luminosity based on 2MASS photometry and bolometric correction based on APOGEE and \emph{Gaia} DR2 parallax & $\mathrm{L_{\odot}}$ \\
\texttt{evstate\_MV} & Evolutionary state using Vrard et al. (2016). RGB = 0, RC = 1, unclassified = -1 & $\mathrm{None}$ \\
\texttt{lum\_flag\_BHM} & Luminosity flag for BHM - 1 if photometric and seismic luminosity agree within $3\sigma_{c}$ & $\mathrm{boolean}$ \\
\texttt{lum\_flag\_COR} & Luminosity flag for COR - 1 if photometric and seismic luminosity agree within $3\sigma_{c}$ & $\mathrm{boolean}$ \\
\texttt{lum\_flag\_A2Z} & Luminosity flag for A2Z- 1 if photometric and seismic luminosity agree within $3\sigma_{c}$ & $\mathrm{boolean}$ \\
\texttt{numax\_dnu\_consistent} & Quality flag, 1 if results on $\nu_{\mathrm{max}}$ and $\Delta\nu$ are consistent within $3\sigma$ across all pipelines & $\mathrm{boolean}$ \\
\texttt{numax\_predicted} & $\nu_{\mathrm{max}}$ predicted using 2MASS photmetry and Gaia DR2 parallax & $\mathrm{\mu Hz}$ \\
\hline
\end{tabular}
\end{table*}

\newpage 
{\small \it $^{1}$School of Astronomy and Astrophysics, University of Birmingham, Edgbaston, Birmimgham, B15 2TT, UK \\
$^{2}$ Canadian Institute for Theoretical Astrophysics, University of Toronto, 60 St. George Street, Toronto, ON, M5S 3H8, Canada \\
$^{3}$ Dunlap Institute for Astronomy and Astrophysics, University of Toronto, 50 St. George Street, Toronto, ON M5S 3H4, Canada \\
$^{4}$ David A. Dunlap Department for Astronomy and Astrophysics, University of Toronto, 50 St. George Street, Toronto, ON M5S 3H4, Canada \\
$^{5}$LESIA, Observatoire de Paris, Universit\'e PSL, CNRS, Sorbonne Universit\'e, Universit\'e de Paris, 5 place Jules Janssen, 92195 Meudon, France \\
$^{6}$Instituto de Astrof\'{\i}sica de Canarias, La Laguna, Tenerife, Spain \\
$^{7}$Dpto. de Astrof\'{\i}sica, Universidad de La Laguna, La Laguna, Tenerife, Spain \\
$^{8}$IRFU, CEA, Universit\'e Paris-Saclay, F-91191 Gif-sur-Yvette, France \\
$^{9}$AIM, CEA, CNRS, Universit\'e Paris-Saclay, Universit\'e Paris Diderot, Sorbonne Paris Cit\'e, F-91191 Gif-sur-Yvette, France \\
$^{10}$Aix Marseille Univ, CNRS, CNES, LAM, Marseille, France \\
$^{11}$Istituto Nazionale Astrofisica di Padova - Osservatorio Astronomico di Padova, Vicolo dell'Osservatorio 5, IT-35122, Padova, Italy \\
$^{12}$Dept. of Astronomy, The Ohio State University, 140 W. 18th Ave., Columbus, OH 43210, USA \\
$^{13}$Department of Astronomy, Yale University, New Haven, CT, 06520, USA \\
$^{14}$Department of Astrophysical Sciences, Princeton University, 4 Ivy Lane, Princeton, NJ~08544 \\
$^{15}$The Observatories of the Carnegie Institution for Science, 813 Santa Barbara St., Pasadena, CA~91101 \\
$^{16}$Institut f\"ur Physik, Universit\"at Graz, Universit\"atsplatz 5/II, 8020 Graz \\
$^{17}$Max-Planck Institut f\"ur Astronomie, K\"onigstuhl 17, 69117 Heidelberg, Germany\\
$^{18}$Instituto de Astrof\'{\i}sica e Ci\^{e}ncias do Espa\c{c}o, Universidade do Porto,  Rua das Estrelas, 4150-762 Porto, Portugal\\
$^{19}$Research School of Astronomy and Astrophysics, Mount Stromlo Observatory, The Australian National University, ACT 2611, Australia\\
$^{20}$Departamento de F\'{\i}sica e Astronomia, Faculdade de Ci\^{e}ncias da Universidade do Porto, Rua do Campo Alegre, s/n, 4169-007 Porto, Portugal\\
$^{21}$Leibniz-Institut f\"ur Astrophysik Potsdam (AIP), An der Sternwarte 16, 14482 Potsdam, Germany\\
$^{22}$Department of Astronomy, The Ohio State University, Columbus, OH 43210, USA\\
$^{23}$Institute of Space Sciences (ICE, CSIC), Carrer de Can Magrans S/N, E-08193, Bellaterra, Spain\\
$^{24}$Institut d'Estudis Espacials de Cataluna (IEEC), Carrer Gran Capita 2, E-08034, Barcelona, Spain\\
$^{25}$Stellar Astrophysics Centre, Department of Physics and Astronomy, Aarhus University, Ny Munkegade 120, DK-8000 Aarhus C, Denmark\\
$^{26}$School of Physics, University of New South Wales, NSW 2052, Australia\\
$^{27}$Institute for Astronomy, University of Hawai‘i, 2680 Woodlawn Drive, Honolulu, Hawaii 96822, USA\\
$^{28}$Carnegie Earth and Planet Laboratory, 5241 Broad Branch Road, N.W., Washington, DC 20015\\
$^{29}$The Observatories of the Carnegie Institution for Science, 813 Santa Barbara St., Pasadena, CA~91101\\}

\bsp	
\label{lastpage}
\end{document}